\documentclass[showpacs,amssymb,floatfix,pre,aps,notitlepage,preprint]{revtex4-1}

\usepackage{amssymb,amsfonts,bm}
\usepackage{graphics,graphicx}
\usepackage{colordvi,amsmath,epsfig,color}
\usepackage[activeacute,english]{babel}
\newcommand{\sig}{\boldsymbol{\hat{\sigma}}}

\begin{document}

\title{Fluctuations in the Uniform Shear Flow state of a granular gas}
\author{M. I. Garc\'ia de Soria}
\author{P. Maynar}
\author{J. Javier Brey}
\affiliation{F\'{\i}sica Te\'{o}rica, Universidad de Sevilla,
Apartado de Correos 1065, E-41080, Sevilla, Spain}

\begin{abstract}
We study the fluctuations of the total internal energy of a granular gas under 
stationary uniform shear flow by means of kinetic theory methods. We find that 
these fluctuations are coupled to the fluctuations of the different components 
of the total pressure tensor. Explicit expressions for all the possible cross 
and auto correlations of the fluctuations at one and two times are obtained 
in the two dimensional case. The theoretical predictions are compared with 
Molecular Dynamics simulation and a good agreement is found for the range of 
inelasticity considered. 
\end{abstract}
\maketitle

\section{Introduction}
A granular system can be defined as an ensemble of macroscopic particles 
(grains) that collide inelastically, i.e. kinetic energy is 
dissipated in collisions. When the dynamics of the grains can be partitioned 
into sequences of two-body collisions, the system is referred to a granular 
gas and there is support both from experiments and 
computer simulations of the reliability of a kinetic theory description 
\cite{d00,g03,bplibro04,l05,at06}. One of the most used models to study 
granular gases is the Inelastic 
Hard Sphere (IHS) model, whose dynamics is given in terms of free streaming 
followed by 
instantaneous inelastic collisions. For this model, all the kinetic theory 
machinery can be applied \cite{resibois}. In particular, in the low density 
limit the dynamics of 
the one-particle distribution function is given by the inelastic Boltzmann 
equation \cite{gs95,bds97} and the correlation functions obey a closed set of 
equations \cite{bgmr04}. 

Macroscopically, it is known that, in many cases, the 
dynamics of a granular system is reminiscent of that of a fluid. For dilute 
systems, hydrodynamic equations can be derived applying the 
Chapman-Enskog expansion \cite{bdks98} or linear response methods 
\cite{bdr03,db03,bd05}, obtaining explicit expressions for the transport 
coefficients. In all these studies, there is a particular state which plays a 
specially important role; the Homogeneous Cooling State (HCS). This is a 
homogeneous state in which all the time dependence in the one-particle 
distribution function 
goes through the granular temperature (defined as the second velocity moment of 
the velocity distribution). Due to the inelasticity of collisions, the 
temperature 
decays monotonically in time \cite{h83}. It is known that, for a wide class of 
initial conditions, the HCS is reached in the long-time limit for isolated 
granular gases. This fact makes that this state play, for granular gases, a 
similar role to the equilibrium state in the context of molecular, elastic 
fluids. In fact, the zeroth order in the gradients distribution in the 
Chapman-Enskog expansion is a ``local'' HCS \cite{bdks98}. 

Despite this analogy with normal fluids, there is also important differences. 
Due to the macroscopic character of the grains, a granular system contains 
typically much less particles than a normal fluid. This fact makes that the 
fluctuations of the 
macroscopic fields be of special relevance not only theoretically, 
but also from a practical point of view. The fluctuations of the total energy 
have been studied in the HCS, and explicit 
expressions for its variance and two-time correlation function have been 
obtained \cite{bgmr04}. With some generality, Langevin-like equations for the 
fluctuating hydrodynamic fields have been 
derived to Navier-Stokes order \cite{bmg09,bmg11}, finding that there are not 
Fluctuation-Dissipation theorems of the second kind, i.e. the amplitude of the 
noises are not related to the transport coefficients. On the other hand, the 
two-time correlation functions do decay as a macroscopic perturbation so that 
Fluctuation-Dissipation theorems of first kind hold \cite{kth87}. 

The study of the fluctuations in the HCS is of special relevance, because it 
serves as a starting point for the generalization to other states. Making an 
analogy with normal fluids, the equations for the fluctuating fields 
can be written in a intuitive manner for states that are close to the HCS. The 
deterministic part of the equations is the linearization of the macroscopic 
equations around the particular state considered. The noises can be assumed to 
have the same stochastic properties that in the HCS but replacing the total 
fields by the local actual ones. 
As said, this is expected to be valid if the state is not far from the 
HCS, which means small gradients. 
In fact, this idea was applied in \cite{bgm12} to calculate the total internal 
energy fluctuations in the stationary Uniform Shear Flow 
(USF) state. This state is characterized by a uniform density, a constant and 
uniform temperature, and a flow velocity with a linear profile and, due to its 
simplicity, it has been extensively studied 
\cite{lsj84,jr88,sgn96,brm97,GSlibro, sgd04}. 
The theoretical predictions of \cite{bgm12} were expected to hold only for 
small gradients, that for the USF 
means small inelasticity due to the coupling between gradients and 
inelasticity, which is a characteristic feature of stationary states of 
granular systems. The objective of this 
work is the study of the fluctuations of the total internal energy in the 
stationary USF state using kinetic theory tools. This will let us analyze the 
problem in general (without any limitation to small inelasticity) and, in 
particular, the differences with the ``local'' HCS results of \cite{bgm12}. It 
will be shown that the structure of these fluctuations is more 
complex than expected, since they are coupled to the 
fluctuations of the several components of the total pressure tensor. In the 
end, a systematic and controlled expansion in the degree of inelasticity will 
be done, in order to be able to get explicit results. 

The plan of the paper is as follows. In the next section, the IHS model is 
described in some detail, and the evolution equations for the relevant 
distributions 
are summarized. These equations are applied to the stationary USF state in 
section \ref{section3}, where the specific case of correlations of global 
quantities 
is considered. In section \ref{section4} we study the fluctuations of the 
total internal energy and it is shown that they are coupled to the components 
of the 
total pressure tensor as mentioned above. The complete study of all the 
fluctuations is carried out in section \ref{section5}. The analytical 
predictions are compared to Molecular Dynamics 
simulation results in section \ref{section6}, finding, in general, a good 
agreement. The final section contains some general conclusions and comments.

\section{Kinetic equations for the model}\label{section2}

The system we consider is a dilute gas of $N$ smooth inelastic hard spheres 
($d=3$) or disks ($d=2$) of mass $m$ and diameter $\sigma$. Let 
$X_i(t)\equiv\{\mathbf{R}_i(t),\mathbf{V}_i(t)\}$ denote the position and 
velocity of particle $i$ at time $t$. The dynamical state of the system, 
$\Gamma(t)\equiv\{X_1(t),\dots,X_N(t)\}$, is generated by free streaming 
followed by instantaneous inelastic collisions characterized by a coefficient 
of normal restitution, $\alpha$, independent of the relative velocity. 
If at time $t$ there is a binary encounter between particles 
$i$ and $j$, with velocities $\mathbf{V}_i(t)$ and $\mathbf{V}_j(t)$ 
respectively, the postcollisional velocities $\mathbf{V}_i'(t)$ and 
$\mathbf{V}_j'(t)$ are
\begin{eqnarray}\label{collisionRule}
\mathbf{V}_i'&=&\mathbf{V}_i-\frac{1+\alpha}{2}
(\hat{\boldsymbol{\sigma}}\cdot\mathbf{V}_{ij})\hat{\boldsymbol{\sigma}},
\nonumber\\
\mathbf{V}_j'&=&\mathbf{V}_j+\frac{1+\alpha}{2}
(\hat{\boldsymbol{\sigma}}\cdot\mathbf{V}_{ij})\hat{\boldsymbol{\sigma}},
\end{eqnarray}
where $\mathbf{V}_{ij}\equiv\mathbf{V}_i-\mathbf{V}_j$ is the relative
velocity and $\hat{\boldsymbol{\sigma}}$ is the unit vector pointing from the
center of particle $j$ to the center of particle $i$ at contact. 

Microscopic densities in phase space, $F_s(x_1,\dots,x_s,t)$, are defined by
\begin{equation}
F_1(x_1,t)=\sum_{i=1}^N\delta[x_1-X_i(t)], 
\end{equation}
\begin{equation}
F_2(x_1,x_2,t)=\sum_{i=1}^N\sum_{j\ne i}^N\delta[x_1-X_i(t)]\delta[x_2-X_j(t)], 
\end{equation}
etc, where we have introduced the field variables, 
$x_i\equiv\{\mathbf{r}_i, \mathbf{v}_i\}$. 
The averages of the microscopic densities over the probability distribution 
function, $\rho(\Gamma,0)$, characterizing the initial state are the 
usual one-time reduced distribution functions
\begin{equation}
f_s(x_1,\dots,x_s,t)\equiv\langle F_s(x_1,\dots,x_s,t)\rangle, 
\end{equation}
where we have introduced the notation
\begin{equation}
\langle G\rangle\equiv\int d\Gamma G(\Gamma)\rho(\Gamma,0). 
\end{equation}
Two-time reduced distribution functions can also be defined in terms of the 
microscopic densities as 
\begin{equation}\label{2timedist}
f_{r,s}(x_1,\dots,x_r,t;x_1,\dots,x_s,t')\equiv\langle F_r(x_1,\dots,x_r,t)
F_s(x_1,\dots,x_s,t')\rangle, 
\end{equation}
where it will be assumed that $t>t'>0$ for concreteness. Evolution 
equations for the reduced distributions can be derived form first principles 
\cite{bds97,bgmr04}, in 
the same way as in the elastic case \cite{ec81}. The one-time reduced 
distribution functions obey the generalization for inelastic collisions of the 
Bogoliubov, Born, Green, Kirkwood and Yvon hierarchy, but its application in 
general is 
limited due to the fact that the equations are not closed. The same occurs for 
the two-time reduced distribution functions. 

It is convenient to introduce correlation functions through the usual cluster 
expansion. From the one-time reduced distributions, one-time correlations, 
$g_s(x_1,\dots,x_s,t)$, are defined by
\begin{equation}
f_2(x_1,x_2,t)\equiv f_1(x_1,t)f_1(x_2,t)+g_2(x_1,x_2,t), 
\end{equation}
\begin{eqnarray}
f_3(x_1,x_2,x_3,t)\equiv f_1(x_1,t)f_1(x_2,t)f_1(x_3,t)+f_1(x_1,t)g_2(x_2,x_3,t)
\nonumber\\
+f_1(x_2,t)g_2(x_1,x_3,t)+f_1(x_3,t)g_2(x_1,x_2,t)+g_3(x_1,x_2,x_3,t), 
\end{eqnarray}
etc. Similarly, two-time correlations functions, 
$h_{r,s}(x_1,\dots,x_r,t;x_1,\dots,x_s,t')$, can be defined. In particular, 
$h_{1,1}$ and $h_{2,1}$ are introduced through
\begin{equation}\label{defh11}
f_{1,1}(x_1,t;x_1',t')=f_1(x_1,t)f_1(x_1',t')+h_{1,1}(x_1,t;x_1',t'), 
\end{equation}
\begin{eqnarray}
f_{2,1}(x_1,x_2,t;x_1',t')=f_1(x_1,t)f_1(x_2,t)f_1(x_1',t')
+g_2(x_1,x_2,t)f_1(x_1',t')\nonumber\\
+h_{1,1}(x_1,t;x_1',t')f_1(x_2,t)
+h_{1,1}(x_2,t;x_1',t')f_1(x_1,t)+h_{2,1}(x_1,x_2,t;x_1',t'). 
\end{eqnarray}
In the low density limit and for distances much longer than the diameter of 
the particles, a closed set of equations for $f_1$, $g_2$ and $h_{1,1}$ is 
obtained \cite{bgmr04,ec81}. 

The one particle distribution function satisfies the inelastic Boltzmann 
equation \cite{bds97,gs95}
\begin{equation}\label{ecB}
\left[\frac{\partial}{\partial t}+L^{(0)}(x_1)\right]f_1(x_1,t)=J[x_1,t|f_1], 
\end{equation}
where we have introduced the free-streaming operator
\begin{equation}
L^{(0)}(x_1)=\mathbf{v}_1\cdot\frac{\partial}{\partial\mathbf{r}_1}.  
\end{equation}
The collisional term reads
\begin{equation}
J[x_1,t|f_1]=\int dx_2\delta(\mathbf{r}_{12})
\overline{T}_0(\mathbf{v}_1,\mathbf{v}_2)f_1(x_1,t)f_1(x_2,t), 
\end{equation}
with the binary collision operator, $T_0$, given by 
\begin{equation}
\overline{T}_0(\mathbf{v}_1,\mathbf{v}_2)=\sigma^{d-1}\int d\boldsymbol{\hat{\sigma}}
\Theta(\mathbf{v}_{12}\cdot\sig)(\mathbf{v}_{12}\cdot\sig)
[\alpha^{-2}b_{\boldsymbol{\sigma}}^{-1}(1,2)-1].  
\end{equation}
Here $\boldsymbol{\sigma}=\sigma\sig$, $d\sig$ is the solid angle element for 
$\sig$, $\mathbf{v}_{12}\equiv\mathbf{v}_1-\mathbf{v}_2$, $\Theta$ is the 
Heaviside step function and the operator $b_{\boldsymbol{\sigma}}^{-1}(1,2)$ 
replaces all the velocities $\mathbf{v}_1$ and $\mathbf{v}_2$ appearing to its 
right by the precollisional values $\mathbf{v}_1^*$ and $\mathbf{v}_2^*$, 
\begin{eqnarray}\label{invCR}
\mathbf{v}_1^*\equiv b^{-1}_{\boldsymbol{\sigma}}(1,2)\mathbf{v}_1&=&
\mathbf{v}_1-\frac{1+\alpha}{2\alpha}(\sig\cdot\mathbf{v}_{12})\sig,
\nonumber\\
\mathbf{v}_2^*\equiv b^{-1}_{\boldsymbol{\sigma}}(1,2)\mathbf{v}_2&=&
\mathbf{v}_2+\frac{1+\alpha}{2\alpha}(\sig\cdot\mathbf{v}_{12})\sig.
\end{eqnarray}

The equation for the one-time correlation function in the low density limit is
\begin{eqnarray}\label{ecg}
\left[\frac{\partial}{\partial t}+L^{(0)}(x_1)+L^{(0)}(x_2)-K[x_1,t|f_1]
-K[x_2,t|f_1]\right]g_2(x_1,x_2,t)\nonumber\\
=\delta(\mathbf{r}_{12})
\overline{T}_0(\mathbf{v}_1,\mathbf{v}_2)f_1(x_1,t)f_1(x_2,t), 
\end{eqnarray}
where we have introduced the linear operator
\begin{equation}\label{defK}
K[x_i,t|f_1]\equiv\int dx_3\delta(\mathbf{r}_{i3})
\overline{T}_0(\mathbf{v}_i,\mathbf{v}_3)(1+P_{i3})f_1(x_3,t). 
\end{equation}
The operator $P_{ij}$ interchanges the labels of particle $i$ and $j$ in the 
quantities to its right. Basically, Eq. (\ref{ecg}) shows that velocity 
correlations between 
particles with velocities $\mathbf{v}_1$ and $\mathbf{v}_2$ are 
generated by uncorrelated collisions implying particles with velocities 
$\mathbf{v}_1$ and 
$\mathbf{v}_2$ through the right hand side of Eq. (\ref{ecg}), and correlated 
collisions implying two particles with velocities $\mathbf{v}_1$ or 
$\mathbf{v}_2$ and a third particle with velocity $\mathbf{v}_3$ through the 
linear operator $K$. 

Finally, let us consider the two-time correlations. In the same limit, the 
following equation for the first two-time correlation function is obtained 
\begin{equation}\label{ech}
\left[\frac{\partial}{\partial t}+L^{(0)}(x_1)-K[x_1,t|f_1]\right]
h_{1,1}(x_1,t;x_1',t')=0. 
\end{equation}
This equation has to be solved with the initial condition
\begin{equation}\label{ich11}
h_{1,1}(x_1,t';x_1',t')=f_1(x_1,t')\delta(x_1-x_1')+g_2(x_1,x_1',t'), 
\end{equation}
that follows directly from the definitions in Eqs. (\ref{2timedist}) and 
(\ref{defh11}). Let us remark that if we consider states with a one-particle 
distribution function, $\tilde{f}_1$, very 
closed to a given reference distribution, $f_1$, the difference of both 
distributions, $\delta f_1\equiv\tilde{f}_1-f_1$, fulfills to linear order 
\begin{equation}\label{ecDf}
\left[\frac{\partial}{\partial t}+L^{(0)}(x_1)-K[x_1,t|f_1]\right]
\delta f_1(x_1,t)=0,  
\end{equation}
where $K$ is given by Eq. (\ref{defK}). 
The structure of these equations is important because it shows that the 
two-time correlations in a given state decays in the same 
way that a linear perturbation of the one-particle distribution function around 
this state. This is so because the linear operator governing the dynamics is, 
in both cases, the operator $K$ given by Eq. (\ref{defK}). Of course, although 
the initial condition of Eq. (\ref{ecDf}) is free (with the only restriction 
to have $|\delta f_1(x_1,t')|<<f_1(x_1,t')$), the initial condition for Eq. 
(\ref{ech}) is given by Eq. (\ref{ich11}). 

\section{The  stationary Uniform Shear Flow state}\label{section3}

In this section we are going to apply the equations of the previous section to 
a particular state, the stationary USF state. At a 
macroscopic level, 
this state is characterized by a uniform number density, $n_s$, a stationary 
temperature, $T_s$, and a constant velocity field with linear profile, 
$\mathbf{u}_s=ay\mathbf{\hat{e}}_x$, where $a$ is the 
constant shear rate and $\mathbf{\hat{e}}_x$ is a unit vector in the direction 
of the $x$-axis (the subindex $s$ has been 
introduced to label the state) \cite{jr88,sgn96, brm97,GSlibro}. In this 
stationary state, the cooling due to collisions is compensated by viscous 
heating
\begin{equation}
\frac{2a}{dn_s}P_{xy, s}=\zeta_sT_s, 
\end{equation}
where $P_{xy, s}$ is the $xy$ component of the stress tensor and $\zeta_s$ is 
the cooling rate. For a hydrodynamic description, $P_{xy, s}$ and $\zeta_s$ 
have to be expressed in terms of the hydrodynamic fields, $n_s$, $T_s$ and 
$\mathbf{u}_s$, and their gradients, i.e. the shear rate, $a$ 
\cite{bdks98,dblibro06}. 

The USF state can be studied by means of kinetic theory. The 
definitions of the hydrodynamic fields in terms of the one-particle 
distribution function are the usual in kinetic theory
\begin{eqnarray}
n(\mathbf{r},t)&\equiv&\int d\mathbf{v}f_1(x,t), \\
n(\mathbf{r},t)\mathbf{u}(\mathbf{r},t)&\equiv&\int d\mathbf{v}\mathbf{v}
f_1(x,t), \\
\frac{d}{2}n(\mathbf{r},t)T(\mathbf{r},t)&\equiv&
\int d\mathbf{v}\frac{m}{2}[\mathbf{v}-\mathbf{u}(\mathbf{r},t)]^2f_1(x,t). 
\end{eqnarray}
The expressions of the pressure tensor and cooling rate are \cite{bdks98}
\begin{equation}
P_{ij}(\mathbf{r},t)
=m\int d\mathbf{v}[v_i-u_i(\mathbf{r},t)][v_j-u_j(\mathbf{r},t)]f_1(x,t), 
\end{equation}
and 
\begin{equation}
\zeta(\mathbf{r},t)=
\frac{(1-\alpha^2)\pi^{(d-1)/2}m\sigma^{d-1}}{4d\Gamma\left(\frac{d+3}{2}\right)
n(\mathbf{r},t)T(\mathbf{r},t)}
\int d\mathbf{v}_1\int d\mathbf{v}_2v_{12}^3
f_1(\mathbf{r},\mathbf{v}_1,t)f_1(\mathbf{r},\mathbf{v}_2,t). 
\end{equation}
The Boltzmann equation admits a normal solution of the USF type, i.e. a 
solution in which all the space dependence goes through the hydrodynamic fields 
and all their gradients. As the only space-dependent field is linear, the 
distribution function can be written in a particularly simple form
\begin{equation}\label{deffs}
f_{USF}(\mathbf{r},\mathbf{v})
=f_s[\mathbf{v}-\mathbf{u}_s(\mathbf{r}), n_s,T_s,a]. 
\end{equation}
By substituting this expression into the Boltzmann equation, it follows that 
\begin{equation}\label{sbe}
aV_{1y}\frac{\partial}{\partial V_{1x}}f_s(\mathbf{V}_1)+\int d\mathbf{V}_2
\overline{T}_0(\mathbf{V}_1,\mathbf{V}_2)f_s(\mathbf{V}_1)f_s(\mathbf{V}_2)=0, 
\end{equation}
where we have introduced the peculiar velocity 
$\mathbf{V}\equiv\mathbf{v}-\mathbf{u}_s(\mathbf{r})$, and we have skipped the 
explicit dependence on the hydrodynamic fields and the shear rate in the 
distribution function. Note that the form of the distribution given by Eq. 
(\ref{deffs}) implies that the system is homogeneous in the Lagrangian frame 
of reference. Although the exact solution of Eq. (\ref{sbe}) is not 
known, many approximate solutions are available \cite{jr88,sgn96,brm97,sgd04}. 
In this work we will consider the $\epsilon\equiv(1-\alpha^2)^{1/2}$ 
expansion of the Jenkins and Richman approximation up to order $\epsilon^2$. 
The specific form of the distribution will be given later on. 

It is convenient to perform the following change of variables
\begin{equation}
\{\mathbf{r},\mathbf{v},t\}\longrightarrow
\{\boldsymbol{\ell}(\mathbf{r},t)=\mathbf{r}-ayt\mathbf{\hat{e}}_x,
\mathbf{V}(\mathbf{r},t)=\mathbf{v}-ay\mathbf{\hat{e}}_x,t\}. 
\end{equation}
Actually, the USF is usually generated in particle simulations using 
Lees-Edwards boundary conditions \cite{le72} and, as stressed in \cite{dsbr86}, 
these boundary conditions transform into periodic boundary conditions in the 
new variables. As the Jacobian of the transformation is one, the function
\begin{equation}
f(\boldsymbol{\ell},\mathbf{V},t)=f_1[\mathbf{r}(\boldsymbol{\ell},t), 
\mathbf{v}(\boldsymbol{\ell},\mathbf{V},t), t], 
\end{equation}
is the actual distribution function in the new variables. Let us consider 
situations very closed to the USF state, in such a way that the deviations, 
$\delta f(\boldsymbol{\ell},\mathbf{V},t)\equiv 
f(\boldsymbol{\ell},\mathbf{V},t)-f_s(\mathbf{V})$, are assumed to fulfill the 
condition
$|\delta f(\boldsymbol{\ell},\mathbf{V},t)|<<f_s(\mathbf{V})$. To linear order, 
$\delta f$ satisfies
\begin{equation}\label{eqdf}
\frac{\partial}{\partial t}\delta f(\boldsymbol{\ell},\mathbf{V}_1,t)=
H(\boldsymbol{\ell},\mathbf{V}_1,t)\delta f(\boldsymbol{\ell},\mathbf{V}_1,t), 
\end{equation}
where
\begin{equation}\label{opH}
H(\boldsymbol{\ell},\mathbf{V}_1,t)\equiv L(\mathbf{V}_1)
-\mathbf{V}_1\cdot\frac{\partial}{\partial\boldsymbol{\ell}}
-a\ell_{y}\frac{\partial}{\partial \ell_{x}}
+atV_{1y}\frac{\partial}{\partial \ell_{x}} 
\end{equation}
is an inhomogeneous linear operator with
\begin{equation}\label{lop}
L(\mathbf{V}_1)h(\mathbf{V}_1)\equiv\int d\mathbf{V}_2
\overline{T}_0(\mathbf{V}_1,\mathbf{V}_2)(1+P_{12})f_s(\mathbf{V}_1)h(\mathbf{V}_2)
+aV_{1y}\frac{\partial}{\partial V_{1x}}h(\mathbf{V}_1). 
\end{equation}
Note that, in contrast with the free cooling case \cite{bdr03,db03,bd05}, the 
inhomogeneous term in Eq. (\ref{opH}) is time dependent. 

Consider now the one-time and two-time correlation functions in the 
USF, $g_{2 USF}$ and $h_{1,1 USF}$ respectively. In the new variables, the 
equations for
\begin{equation}
G_s(\boldsymbol{\ell}_1,\mathbf{V}_1,\boldsymbol{\ell}_2,\mathbf{V}_2)
\equiv g_{2 USF}(x_1,x_2), 
\end{equation}
and 
\begin{equation}
h_s(\boldsymbol{\ell}_1,\mathbf{V}_1,t;\boldsymbol{\ell}_2,\mathbf{V}_2,t')
\equiv h_{1,1 USF}(x_1,t;x_2,t'), 
\end{equation}
are
\begin{equation}\label{eqGs}
\left[H(\boldsymbol{\ell}_1,\mathbf{V}_1,t)+H(\boldsymbol{\ell}_2,\mathbf{V}_2,t)\right]
G_s(\boldsymbol{\ell}_1,\mathbf{V}_1,\boldsymbol{\ell}_2,\mathbf{V}_2)
=-\delta(\boldsymbol{\ell}_{12})\overline{T}_0(\mathbf{V}_1,\mathbf{V}_2)
f_s(\mathbf{V}_1)f_s(\mathbf{V}_2), 
\end{equation}
and 
\begin{equation}\label{eqhs}
\frac{\partial}{\partial t}
h_s(\boldsymbol{\ell}_1,\mathbf{V}_1,t;\boldsymbol{\ell}_2,\mathbf{V}_2,t')=
H(\boldsymbol{\ell}_1,\mathbf{V}_1,t)
h_s(\boldsymbol{\ell}_1,\mathbf{V}_1,t;\boldsymbol{\ell}_2,\mathbf{V}_2,t'), 
\end{equation}
respectively. This last equation has to be solved with the initial condition 
(see Eq. (\ref{ich11}))
\begin{equation}
h_s(\boldsymbol{\ell}_1,\mathbf{V}_1,t';\boldsymbol{\ell}_2,\mathbf{V}_2,t')=
f_s(\mathbf{V}_1)\delta(\boldsymbol{\ell}_{12})\delta(\mathbf{V}_{12})
+G_s(\boldsymbol{\ell}_1,\mathbf{V}_1,\boldsymbol{\ell}_2,\mathbf{V}_2). 
\end{equation}
Equations (\ref{eqGs}) and (\ref{eqhs}) describe two particle correlations at 
one and two times. Basically, they depend on the one-particle distribution 
function, which is supposed to be known, and on the linear operator defined by 
Eq. (\ref{lop}). 

\subsection{Global correlations}

As the general problem is quite involved, in the following we will focus on a 
simplified problem: the study of the correlations between global quantities. 
In order to deal with dimensionless distributions, we introduce 
the dimensionless velocity
\begin{equation}
\mathbf{c}=\frac{\mathbf{V}}{v_s}, \qquad v_s=\sqrt{\frac{2T_s}{m}}, 
\end{equation}
through the thermal velocity in the stationary USF, $v_s$, and the 
dimensionless time
\begin{equation}
s=\frac{v_s}{\lambda}t, \qquad \lambda=(n_s\sigma^{d-1}), 
\end{equation}
with $\lambda$ proportional to the mean free path. In terms of these units, we 
define the dimensionless distributions. The scaled one-particle distribution 
function in the USF is 
\begin{equation}\label{defchi}
\chi(\mathbf{c})\equiv\frac{v_s^d}{n_s}f_s(\mathbf{V}), 
\end{equation}
the integrated deviation of the one-particle distribution function around the 
USF is 
\begin{equation}
\delta\chi(\mathbf{c},s)\equiv\frac{v_s^d}{n_s}\int d\boldsymbol{\ell}
\delta f(\boldsymbol{\ell}, \mathbf{V},t),  
\end{equation}
the dimensionless marginal one-time correlation function is 
\begin{equation}\label{defphi}
\phi(\mathbf{c}_1,\mathbf{c}_2)\equiv\frac{v_s^{2d}}{N}
\int d\boldsymbol{\ell}_1\int d\boldsymbol{\ell}_2
G_s(\boldsymbol{\ell}_1,\mathbf{V}_1,\boldsymbol{\ell}_2,\mathbf{V}_2), 
\end{equation}
and the dimensionless marginal two-time correlation function is 
\begin{equation}\label{defpsi}
\psi(\mathbf{c}_1,\mathbf{c}_2,s-s')\equiv\frac{v_s^{2d}}{N}
\int d\boldsymbol{\ell}_1\int d\boldsymbol{\ell}_2
h_s(\boldsymbol{\ell}_1,\mathbf{V}_1,t;\boldsymbol{\ell}_2,\mathbf{V}_2,t'). 
\end{equation}

For homogeneous states in the Lagrangian frame of reference, the evolution 
equation for $\delta\chi$, obtained by integrating of Eq. (\ref{eqdf}), reads 
\begin{equation}\label{hbe}
\frac{\partial}{\partial s}\delta\chi(\mathbf{c},s)=
\Lambda(\mathbf{c})\delta\chi(\mathbf{c},s). 
\end{equation}
The operator $\Lambda$ will be called linearized Boltzmann operator and is the 
adimensionalization of the linear operator defined in Eq. (\ref{lop}), i.e., 
\begin{equation}\label{lbco}
\Lambda(\mathbf{c}_1)h(\mathbf{c}_1)\equiv\int d\mathbf{c}_2
\widetilde{T}_0(\mathbf{c}_1,\mathbf{c}_2)(1+P_{12})\chi(\mathbf{c}_1)h(\mathbf{c}_2)
+\tilde{a}_sc_{1y}\frac{\partial}{\partial c_{1x}}h(\mathbf{c}_1), 
\end{equation}
where $\widetilde{T}_0$ is the dimensionless counterpart of $\overline{T}_0$
\begin{equation}
\widetilde{T}_0(\mathbf{c}_1,\mathbf{c}_2)\equiv\int d\sig\Theta(\mathbf{c}_{12}\cdot\sig)
(\mathbf{c}_{12}\cdot\sig)[\alpha^{-2}b_{\boldsymbol{\sigma}}^{-1}(1,2)-1], 
\end{equation}
and 
\begin{equation}
\tilde{a}_s=\frac{\lambda a}{v_s}, 
\end{equation}
is the dimensionless shear rate. 

The equation for the one-time correlation function is
\begin{equation}\label{eqphi}
[\Lambda(\mathbf{c}_1)+\Lambda(\mathbf{c}_2)]\phi(\mathbf{c}_1,\mathbf{c}_2)
=-\widetilde{T}_0(\mathbf{c}_1,\mathbf{c}_2)\chi(\mathbf{c}_1)\chi(\mathbf{c}_2),  
\end{equation}
and the evolution equation for the two-time correlation functions is
\begin{equation}\label{eqPsi}
\frac{\partial}{\partial s}\psi(\mathbf{c}_1,\mathbf{c}_2,s)=
\Lambda(\mathbf{c}_1)\psi(\mathbf{c}_1,\mathbf{c}_2,s), 
\end{equation}
to be solved with the initial condition 
\begin{equation}\label{cipsi}
\psi(\mathbf{c}_1,\mathbf{c}_2,0)=\chi(\mathbf{c}_1)\delta(\mathbf{c}_{12})
+\phi(\mathbf{c}_1,\mathbf{c}_2). 
\end{equation}
It is important to remark the strong analogy between equations (\ref{eqphi}) 
and (\ref{eqPsi}), and the equivalent ones in other granular states (as the 
homogeneous cooling state \cite{bgmr04}), or other granular systems in which 
the particles are accelerated by a stochastic force \cite{gmt09}. The analogy 
is also evident with other 
dissipative systems \cite{mgsbt08}, where the linearized Boltzmann collision 
operator, $\Lambda(\mathbf{c})$, always plays an essential role in the 
structure of the global correlations.

\subsection{Correlations between global quantities}

The correlations between global quantities can be evaluated using the 
distributions we have introduced above. Consider quantities of the form
\begin{equation}\label{flucA}
\mathcal{A}(t)=\sum_{i=1}^Na[\mathbf{V}_i-\mathbf{u}_s(\mathbf{R}_i)]
=\int d\mathbf{r}\int d\mathbf{v}a[\mathbf{v}-\mathbf{u}_s(\mathbf{r})]
F_1(x,t), 
\end{equation}
where $a$ is supposed to be a homogeneous function of degree $\beta$, i.e. 
$a(k\mathbf{c})=k^\beta a(\mathbf{c})$. The deviation around the mean in the 
USF is
\begin{equation}\label{dA}
\delta\mathcal{A}(t)\equiv \mathcal{A}(t)-\langle \mathcal{A}(t)\rangle
=\int d\mathbf{r}\int d\mathbf{v}a(\mathbf{V})\delta F(x,t), 
\end{equation}
where
\begin{equation}
\delta F(x,t)\equiv F_1(x,t)-f_{USF}(x). 
\end{equation}
The correlations between the fluctuations of two different quantities, 
$\mathcal{A}_1$ and $\mathcal{A}_2$, of the form given in Eq. (\ref{flucA}) can 
be expressed as
\begin{eqnarray}\label{a1a2primero}
\langle\delta\mathcal{A}_1(t)\delta \mathcal{A}_2(t')\rangle
=\int d\mathbf{r}_1\int d\mathbf{v}_1\int d\mathbf{r}_2\int d\mathbf{v}_2
a_1(\mathbf{V}_1)a_2(\mathbf{V}_2)h_{1,1 USF}(x_1,t;x_2,t'). 
\end{eqnarray}
Upon writing this expression, we have used that
\begin{equation}
\langle\delta F(x_1,t)\delta F(x_2,t')\rangle
=h_{1,1 USF}(x_1,t;x_2,t'). 
\end{equation}
Expressing the integrand of (\ref{a1a2primero}) in term of the dimensionless 
distribution defined in Eq. (\ref{defpsi}), yields 
\begin{equation}\label{a1a2t}
\langle\delta\mathcal{A}_1(t)\delta\mathcal{A}_2(t')\rangle=Nv_s^{\beta_1+\beta_2}
\int d\mathbf{c}_1\int d\mathbf{c}_2a_1(\mathbf{c}_1)a_2(\mathbf{c}_2)
\psi(\mathbf{c}_1,\mathbf{c}_2,s-s'), 
\end{equation}
where $\beta_1$ and $\beta_2$ are the degree of homogeneity of $a_1$ and $a_2$, 
respectively. 

The expression for the one-time correlations is obtained by performing 
$s=s'$ in Eq. (\ref{a1a2t}). Taking into account Eq. (\ref{cipsi}), it follows 
that 
\begin{equation}\label{a1a2}
\langle\delta \mathcal{A}_1(t)\delta\mathcal{A}_2(t)\rangle=Nv_s^{\beta_1+\beta_2}
\left[\int d\mathbf{c}a_1(\mathbf{c})a_2(\mathbf{c})\chi(\mathbf{c})+
\int d\mathbf{c}_1\int d\mathbf{c}_2a_1(\mathbf{c}_1)a_2(\mathbf{c}_2)
\phi(\mathbf{c}_1,\mathbf{c}_2)\right].  
\end{equation}

\section{Fluctuations of the total internal energy}\label{section4}

Microscopically, the total internal energy is defined as
\begin{equation}
\mathcal{E}(t)
=\sum_{i=1}^N\frac{m}{2}[\mathbf{V}_i-\mathbf{u}_s(\mathbf{R}_i)]^2, 
\end{equation}
so that it is a quantity of the form introduced in the previous section. We 
identify, $a(\mathbf{V})\equiv\frac{m}{2}V^2$, that is a homogeneous function 
of degree two. Using Eq. (\ref{a1a2}), we get 
\begin{equation}\label{flucE}
\langle\delta\mathcal{E}^2(t)\rangle=\frac{m^2}{4}Nv_s^4
\left[\int d\mathbf{c}c^4\chi(\mathbf{c})
+\int d\mathbf{c}_1\int d\mathbf{c}_2c_1^2c_2^2\phi(\mathbf{c}_1,\mathbf{c}_2)
\right]. 
\end{equation}
Since the distribution $\chi$ is supposed to be known, we only have to evaluate 
the velocity moment of $\phi$ that appears in the right hand side of Eq. 
(\ref{flucE}). In order to do that we follow a method based on the analysis of 
some spectral properties of the linearized Boltzmann collision operator, 
$\Lambda$ \cite{bgmr04,mgsbt08,gmt09}. 

\subsection{Spectral properties of $\Lambda$}\label{spectralprop}

In order to identify some modes of the $\Lambda$ operator, it is necessary to 
introduce the time-dependent USF state \cite{as07,as12}. As the 
stationary USF, the time-dependent state is characterized macroscopically by a 
constant density, $n_H$, and a time-independent flow velocity, 
$\mathbf{u}_H(\mathbf{r})=ay\mathbf{\hat{e}_x}$. The temperature, $T_H(t)$, 
remains homogeneous, but it is time-dependent. The subindex $H$ distinguishes 
it from the stationary state labeled by $s$. By dimensional analysis, if a 
normal distribution function for this state exists, it has the form
\begin{equation}\label{tdUSF}
f_H(\mathbf{V},t)=\frac{n_H}{v_H(t)^d}\chi(\mathbf{c},\tilde{a}), 
\end{equation}
where 
\begin{equation}
\mathbf{c}=\frac{\mathbf{V}}{v_H(t)}, \quad v_H(t)=\sqrt{\frac{2T_H(t)}{m}}, 
\quad\tilde{a}=\frac{\lambda a}{v_H(t)}. 
\end{equation}
We use the same notation for the time-dependent scaled velocity 
$\mathbf{V}/v_H(t)$  and for $\mathbf{V}/v_s$ but this will not cause any 
difficulty. In the long time limit, this distribution tends to the stationary 
one
\begin{equation}
\chi(\mathbf{c},\tilde{a})\to\chi(\mathbf{c},\tilde{a}_s)
\equiv\chi(\mathbf{c}),  
\end{equation}
and also the quantities $v_H(t)$ and $\tilde{a}$ to their stationary values 
$v_s$ and $\tilde{a}_s$ respectively. 

Let us consider the family of states given by Eq. (\ref{tdUSF}) with the 
restriction of being close to the stationary USF state. These states are 
characterized by the two parameters
\begin{equation}
\rho\equiv\frac{\delta n}{n_s}, \qquad \theta\equiv\frac{\delta T}{T_s}. 
\end{equation}
It is assumed that the deviations
\begin{equation}
\delta n\equiv n_H-n_s, \quad \delta T\equiv T_H-T_s, 
\end{equation}
are small, i.e. $|\delta n|<<n_s$ and $|\delta T|<<T_s$. We 
do not include states with different shear rates, $a$, because we want all the 
states to be generated 
by the same boundary conditions. Performing a similar analysis to the one 
carried out in reference \cite{bmg12}, the following evolution equation for 
$\theta$
\begin{equation}\label{ectheta}
\frac{d\theta(s)}{ds}=-\gamma[2\rho+\theta(s)], 
\end{equation}
is obtained in Appendix \ref{apendiceA}. 
As the total number of particle does not vary, $\rho$ is constant and we can 
identify the normal mode $[2\rho+\theta(s)]$. The eigenvalue
\begin{equation}\label{defg}
\gamma=\frac{\tilde{\zeta}(\tilde{a}_s)}{2}
-\frac{\tilde{a}_s^2}{d}\frac{d\tilde{P}_{xy}}{d\tilde{a}}(\tilde{a}_s)
-\frac{\tilde{a}_s}{2}\frac{d\tilde{\zeta}}{d\tilde{a}}(\tilde{a}_s), 
\end{equation}
is expressed in terms of the dimensionless pressure tensor
\begin{equation}\label{pxys}
\tilde{P}_{xy}(\tilde{a})=2\int d\mathbf{c}c_xc_y\chi(\mathbf{c},\tilde{a}), 
\end{equation}
and the dimensionless cooling rate
\begin{equation}\label{zetas}
\tilde{\zeta}(\tilde{a})
=\frac{\pi^{(d-1)/2}(1-\alpha^2)}{2d\Gamma\left(\frac{d+3}{2}\right)}
\int d\mathbf{c}_1\int d\mathbf{c}_2c_{12}^3
\chi(\mathbf{c}_1,\tilde{a})\chi(\mathbf{c}_2,\tilde{a}), 
\end{equation}
in the time-dependent USF state. Eq. (\ref {defg}) is equivalent to the one 
derived in \cite{g06} and also to the one of \cite{bmg12} for the case 
$\rho=0$. An explicit 
formula for $\gamma$ as a function of the inelasticity can be written using 
the expressions of $\tilde{P}_{xy}$ and $\tilde{\zeta}(\tilde{a}_s)$ of the BGK 
model studied in \cite{as12} and neglecting the contribution proportional to 
$d\tilde{\zeta}/d\tilde{a}$ . 

\subsection{An unaccurate approximation}

Let us rewrite Eq. (\ref{ectheta}) in a way that suggests the 
approximation to be analyzed in the following. Define the scalar product
\begin{equation}
\langle h(\mathbf{c})|g(\mathbf{c})\rangle
\equiv\int d\mathbf{c}h^*(\mathbf{c})g(\mathbf{c}), 
\end{equation}
the start denoting complex conjugated. 
The deviations $\rho$ and $\theta$ can be expressed in terms of $\delta\chi$ as
\begin{equation}\label{rhotheta}
\rho =\int d\mathbf{c}\delta\chi(\mathbf{c},s), \qquad
\theta(s)=\int d\mathbf{c}\left(\frac{2}{d}c^2-1\right)\delta\chi(\mathbf{c},s),
\end{equation}
and then
\begin{equation}
2\rho+\theta(s)
=\int d\mathbf{c}\left(\frac{2}{d}c^2+1\right)\delta\chi(\mathbf{c},s)
\equiv \langle\bar{\xi}_2(c)|\delta\chi(\mathbf{c},s)\rangle, 
\end{equation}
where we have introduced 
\begin{equation}
\bar{\xi}_2(c)\equiv\frac{2}{d}c^2+1. 
\end{equation}
By taking the scalar product with $\bar{\xi}_2$ in Eq. (\ref{hbe}), it is 
obtained 
\begin{equation}
\frac{d}{ds}\langle\bar{\xi}_2(c)|\delta\chi(\mathbf{c},s)\rangle
=\langle\bar{\xi}_2(c)|\Lambda(\mathbf{c})\delta\chi(\mathbf{c},s)\rangle. 
\end{equation}
Comparing this equation with the evolution equation for $\theta$, Eq. 
(\ref{ectheta}), it is seen 
that, for $\delta\chi$ belonging to the biparametric family of functions of 
time-dependent USF states that are closed to the stationary USF state,
$\langle\bar{\xi}_2(c)|\Lambda(\mathbf{c})\delta\chi(\mathbf{c},s)\rangle 
=-\gamma\langle\bar{\xi}_2(\mathbf{c})|\delta\chi(\mathbf{c},s)\rangle $. 
Below it will be discussed while it is consistent to consider the approximation
\begin{equation}\label{aprox1}
\langle\bar{\xi}_2(c)|\Lambda(\mathbf{c})g(\mathbf{c})\rangle
\approx-\gamma\langle\bar{\xi}_2(\mathbf{c})|g(\mathbf{c})\rangle, 
\end{equation}
for any function, $g(\mathbf{c})$. This is, basically, the approximation that 
allows calculating the fluctuations of the total energy in 
\cite{bgmr04,mgsbt08,gmt09}. Let us also mention that, in the free-cooling 
case, the equivalent of Eq. (\ref{aprox1}) is an exact property for Maxwell 
molecules \cite{bgm10}.

Actually, it will be shown that Eq. (\ref{aprox1}), although consistent 
with linear hydrodynamics, is not consistent with the equation for $\phi$, 
Eq. (\ref{eqphi}). To start with, let us see that some velocity moments of 
$\phi$ can be exactly related to velocity moments of the one-particle 
distribution, $\chi$. As the total number of particles, $N$, does not 
fluctuate, it is evident that 
\begin{equation}
\langle\delta N(t)\delta\mathcal{A}(t)\rangle=0, 
\end{equation}
for any fluctuating quantity, $\mathcal{A}$. If , in addition, $\mathcal{A}$ 
can be expressed as in Eq. (\ref{flucA}), we have
\begin{equation}
\langle\delta N\delta\mathcal{A}(t)\rangle=Nv_s^{\beta}
\left[\int d\mathbf{c}a(\mathbf{c})\chi(\mathbf{c})+
\int d\mathbf{c}_1\int d\mathbf{c}_2a(\mathbf{c}_1)
\phi(\mathbf{c}_1,\mathbf{c}_2)\right], 
\end{equation}
and it can be concluded that
\begin{equation}\label{phiVchi}
\int d\mathbf{c}_1\int d\mathbf{c}_2
a(\mathbf{c}_1)\phi(\mathbf{c}_1,\mathbf{c}_2)
=-\int d\mathbf{c}a(\mathbf{c})\chi(\mathbf{c}), 
\end{equation}
for any homogeneous function, $a(\mathbf{c})$, of degree $\beta$. With this 
property we can easily calculate the component
\begin{equation}\label{p1}
\langle\bar{\xi}_2(c_1)|\phi(\mathbf{c}_1,\mathbf{c}_2)\rangle
=-\int d\mathbf{c}\left(\frac{2c^2}{d}+1\right)\chi(\mathbf{c})=-2. 
\end{equation}
On the other hand, the integral can also be evaluated by taking the scalar 
product with $\bar{\xi}_2$ in the equation for $\phi$, Eq. (\ref{eqphi}), 
obtaining
\begin{equation}\label{p2}
\langle\bar{\xi}_2(c_1)|\phi(\mathbf{c}_1,\mathbf{c}_2)\rangle
=\frac{1}{\gamma}\int d\mathbf{c}_1\int d\mathbf{c}_2
\left(\frac{2c^2}{d}+1\right)\widetilde{T}_0(\mathbf{c}_1,\mathbf{c}_2)
\chi(\mathbf{c}_1)\chi(\mathbf{c}_2)
=-\frac{\tilde{\zeta}_s}{\gamma}, 
\end{equation}
where the expression of $\tilde{\zeta}$, Eq. (\ref{zetas}), has been used, and 
we have introduced the notation 
$\tilde{\zeta}_s\equiv\tilde{\zeta}(\tilde{a}_s)$. Then, it follows that the 
approximation (\ref{aprox1}) is not consistent with the equation for the 
correlation function, Eq. (\ref{eqphi}), because it predicts a different 
result for $\langle\bar{\xi}_2(c_1)|\phi(\mathbf{c}_1,\mathbf{c}_2)\rangle$ 
than the exact one giving by Eq. (\ref{p1}). Moreover, the approximation is not 
even valid in the elastic limit since $\gamma\sim\tilde{\zeta}$ in that limit. 
In fact, when $\langle\delta E^2\rangle$ is calculated using the approximate 
expression of $\langle c_1^2c_2^2|\phi(\mathbf{c}_1,\mathbf{c}_2)\rangle$ 
(evaluated using Eq. (\ref{aprox1})), the obtained result does not agree with 
the results 
of \cite{bgm12} even for $\alpha\to 1$. Of course, this is not surprising, 
since the approximation is not valid in the elastic limit either. 

\subsection {A consistent approximation}

The previous result is clearly unsatisfactory, and it would be desirable to 
find a kind of approximation that would be consistent with both linear 
hydrodynamics and the equation for the correlation function. Let us also note 
that the linearized Boltzmann operator, $\Lambda$, given by Eq. (\ref{lbco}), 
contains a term of the form $c_y\partial/\partial c_x$ which 
mixes the subspace generated by $c^2$ with $c_xc_y$. Then, it can not be 
expected 
that $\langle\bar{\xi}_2|\Lambda\approx -\gamma\langle\bar{\xi}_2|$ be a good 
approximation in general. In fact, the operator $c_y\partial/\partial c_x$ 
leaves invariant the $4$-dimensional subspace generated by 
$\{1,c^2,c_xc_y,c_y^2\}$ and, for Maxwell molecules, the left eigenfunctions of 
$\Lambda$ are linear combination of these $4$ functions \cite{gm}. 
With this in mind, we will search a generalization of approximation 
(\ref{aprox1}) taking as a possible candidate for $\bar{\xi}_2$ a function in 
the subspace generated by $\{1,c^2,c_xc_y,c_y^2\}$. To identify it, we consider 
the evolution equations for the homogeneous pressure tensor components of 
references \cite{brm97,sgd04}
\begin{eqnarray}\label{ceq1}
\frac{\partial T_H}{\partial t}+\zeta_H T_H+\frac{2a}{dn_H}P_{xy, H}=0, \\
\frac{\partial P_{xy, H}}{\partial t}+(\beta\nu_H+\zeta_H)P_{xy, H}+aP_{yy, H}=0,
\label{ceq2} 
\\
\frac{\partial P_{yy, H}}{\partial t}+(\beta\nu_H+\zeta_H)P_{yy, H}
-\beta n_H\nu_H T_H=0,
\label{ceq3} 
\end{eqnarray}
where we have introduced the subindex $H$ to remark that we are only 
considering homogeneous situations. The cooling rate can be expressed as 
\begin{equation}
\zeta_H=\frac{v_H}{\lambda}\tilde{\zeta}_{s}, \quad 
\tilde{\zeta}_s=\frac{\sqrt{2}\pi^{(d-1)/2}(1-\alpha^2)}{d\Gamma(d/2)}, 
\end{equation}
where $\tilde{\zeta}_s$ coincides with $\tilde{\zeta}(\tilde{a}_s)$ calculated 
in the Jenkins and Richman approximation to $\epsilon^2$ order, $\nu_H$ is the 
collision frequency
\begin{displaymath}
\nu_H=\frac{v_H}{\lambda}z, \quad z=\frac{8\pi^{(d-1)/2}}{\sqrt{2}(d+2)\Gamma(d/2)},
\end{displaymath}
and $\beta$ is a parameter to be specified later on.

Equations (\ref{ceq1})-(\ref{ceq3}) admit a stationary solution. Defining the 
dimensionless components of the pressure tensor in the stationary state
\begin{equation}\label{defPijadimen}
\tilde{P}_{ij, s}\equiv\frac{P_{ij, s}}{n_sT_s },  
\end{equation}
it is obtained \cite{brm97,sgd04}
\begin{equation}
\tilde{P}_{xy, s}=-\frac{d\tilde{\zeta_s}}{2\tilde{a}_s}, \quad
\tilde{P}_{yy, s}=\frac{\beta}{\beta+\frac{\tilde{\zeta}_s}{z}}. 
\end{equation}
The dimensionless shear rate is 
\begin{equation}
\tilde{a}_s=z\sqrt{\frac{d\tilde{\zeta}_s}{2 z\beta}}
\left(\beta+\frac{\tilde{\zeta}_s}{z}\right), 
\end{equation}
from which the stationary temperature can be evaluated through 
$v_s=\lambda a /\tilde{a}_s$. Note that all the expressions can be expressed 
in terms of $\tilde{\zeta}_s$ and $\beta$. 

The set of equations (\ref{ceq1})-(\ref{ceq3}), plus the equation for the total 
density that is trivial, can be linearized around the 
stationary state characterized by $n_s$, $T_s$ and $P_{ij, s}$. Defining the 
dimensionless deviations of the pressure tensor as
\begin{equation}
\Pi_{ij}=\frac{P_{ij}-P_{ij, s}}{n_sT_s}, 
\end{equation}
we obtain the following set of linear equations
\begin{equation}\label{eqy}
\frac{d}{\partial s}\mathbf{y}(s)+M\mathbf{y}(s)=\mathbf{0}, 
\end{equation}
for
\begin{equation}\label{eqTC}
\mathbf{y}=\left[\begin{array}{c} 
\rho(s)\\ \theta(s)\\ \Pi_{xy}(s)\\ \Pi_{yy}(s) \end{array}\right], 
\end{equation}
where we have introduced the matrix
\begin{equation}
M=
\left[\begin{array}{cccc}
0 & 0 & 0 & 0\\
2\tilde{\zeta_s} & \frac{3}{2}\tilde{\zeta}_s 
& (\beta z+\tilde{\zeta}_s)\sqrt{\frac{2\tilde{\zeta}_s}{d\beta z}} & 0\\
-\sqrt{\frac{d\tilde{\zeta_s}\beta z}{2}} 
& -\frac{1}{2}\sqrt{\frac{d\tilde{\zeta_s}\beta z}{2}} 
& \tilde{\zeta}_s+\beta z 
& (\beta z+\tilde{\zeta}_s)\sqrt{\frac{d\tilde{\zeta}_s}{2\beta z}} \\
-\beta z & -\beta z & 0 & \tilde{\zeta}_s+\beta z \end{array}\right],  
\end{equation}
that, again, is expressed in terms of $\tilde{\zeta}_s$ and $\beta$. Taking the 
explicit value of $\beta$ evaluated in Grad's approximation \cite{sgd04}
\begin{equation}
\beta=\frac{1+\alpha}{2}\left[1-\frac{d-1}{2d}(1-\alpha)\right],  
\end{equation}
the matrix is expressed uniquely in terms of the inelasticity, $\alpha$. In 
this way, Eq. (\ref{eqy}) is a set of linear differential equations for the 
deviations, $\mathbf{y}$, defined in Eq. (\ref{eqTC}), where  all the 
coefficients of the matrix $M$ are known functions of the coefficient of normal 
restitution, $\alpha$. This is the generalization 
of Eq. (\ref{ectheta}) that we were looking for. The eigenvalues, 
$\{\lambda_i\}_{i=1}^4$, and their corresponding left eigenfunctions of $M$, 
$\{\mathbf{v}_i\}_{i=1}^4$, fulfill
\begin{equation}
\mathbf{v}_i\cdot M=\lambda_i\mathbf{v}_i, 
\end{equation}
and can be calculated with Mathematica. As the expressions 
are very long, here we just write the expansion to $\epsilon^4$ order for $d=2$
\begin{equation}
\lambda_1=0, \quad\lambda_2\approx\sqrt{\frac{\pi}{2}}\epsilon^2
-\frac{3}{4}\sqrt{\frac{\pi}{2}}\epsilon^4, 
\end{equation}
\begin{equation}
\lambda_3\approx
\left(\sqrt{2\pi}+\frac{1}{2}\sqrt{\frac{\pi}{2}}\epsilon^2
+\frac{1}{4}\sqrt{\frac{\pi}{2}}\epsilon^4\right)
-\imath\left(\sqrt{\pi}\epsilon+\frac{19}{64}\epsilon^3\right), 
\quad \lambda_4=\lambda_3^*. 
\end{equation}
The corresponding left eigenfunctions to the same order are
\begin{eqnarray}
\mathbf{v}_1&=&(1,0,0,0), \\
\mathbf{v}_2&\approx&\left(4-\frac{7\epsilon^2}{2}+\frac{17\epsilon^4}{4}, 
2-\frac{11\epsilon^2}{4}+\frac{23\epsilon^4}{8}, 
-\sqrt{2}\epsilon+\frac{11\epsilon^3}{8\sqrt{2}}, \epsilon^2\right), \\
\mathbf{v}_3&\approx&\left(-1-\frac{\epsilon^2}{2}+\epsilon^4, 
-1+\frac{\epsilon^2}{16}+\frac{67\epsilon^4}{128}, 
\frac{\epsilon}{4\sqrt{2}}+\frac{19\epsilon^3}{64\sqrt{2}}, 1\right)\nonumber\\
&-&\imath\left(\frac{3\epsilon^3}{2\sqrt{2}}, 
\frac{\epsilon}{2\sqrt{2}}+\frac{75\epsilon^3}{128\sqrt{2}}, 
1-\frac{\epsilon^2}{64}-\frac{1841\epsilon^4}{8192},0\right),
\\
\mathbf{v}_4&=&\mathbf{v}_3^*. 
\end{eqnarray}
Let us remark that, as Eq. (\ref{ceq1}) was the starting point for the 
derivation of Eq. (\ref{ectheta}), $\lambda_2$ can be expressed in a way 
similar to $\gamma$, 
\begin{equation}
\lambda_2=\frac{\tilde{\zeta}(\tilde{a}_s)}{2}
-\frac{\tilde{a}_s^2}{d}\frac{d\tilde{P}_{xy}}{d\tilde{a}}(\tilde{a}_s).
\end{equation}
Here we do not have the $\frac{d\tilde{\zeta}}{d\tilde{a}}$ contribution since 
it was neglected from the very beginning. 

With the aid of the left eigenfunctions, the normal modes of Eq. (\ref{eqy}) 
can be easily written as
\begin{equation}
\Xi_j
=\mathbf{v}_j\cdot\mathbf{y}=v_{j1}\rho+v_{j2}\theta+v_{j3}\Pi_{xy}+v_{j4}\Pi_{yy}, 
\end{equation}
where $v_{ji}$ is the $i$-th component of $\mathbf{v}_j$. Now, we can identify 
the functions, $\{\bar{\xi}_i(\mathbf{c})\}_{i=1}^4$, 
\begin{equation}\label{defxi}
\bar{\xi}_i(\mathbf{c})=\xi_{i1}+\xi_{i2}c^2+\xi_{i3}c_xc_y+\xi_{i4}c_y^2, 
\end{equation}
such that
\begin{equation}
\langle\bar{\xi}_j(\mathbf{c})|\delta\chi(\mathbf{c})\rangle=\Xi_j. 
\end{equation}
Taking into account Eq. (\ref{rhotheta}) and
\begin{equation}
\Pi_{ij}(s)=2\int d\mathbf{c}c_ic_j\delta\chi(\mathbf{c},s), 
\end{equation}
we can identify
\begin{eqnarray}\label{barxiv1}
\bar{\xi}_1(c)&=&1, \\
\label{barxiv2}
\bar{\xi}_2(\mathbf{c})&=&(v_{21}-v_{22})+\frac{2}{d}v_{22}c^2+2v_{23}c_xc_y
+2v_{24}c_y^2, \\
\label{barxiv3}
\bar{\xi}_3(\mathbf{c})&=&(v_{31}-v_{32})+\frac{2}{d}v_{32}c^2+2v_{33}c_xc_y
+2c_y^2,  \\
\label{barxiv4}
\bar{\xi}_4(\mathbf{c})&=&\bar{\xi}_3^*(\mathbf{c}). 
\end{eqnarray}
Note that while the coefficients $\{v_{2j}\}_{j=1}^4$ are real, 
$\{v_{3j}\}_{j=1}^3$ have an imaginary part. Then the real and imaginary part of 
$\lambda_3$ and $\bar{\xi}_3$ are introduced through
\begin{eqnarray}
\lambda_3&=&\lambda_3^R+\imath\lambda_3^I, \label{lri}\\
\bar{\xi}_3(\mathbf{c})&=&
\bar{\xi}_3^R(\mathbf{c})+\imath\bar{\xi}_3^I(\mathbf{c}). \label{xiri}
\end{eqnarray}

In Appendix \ref{apendiceB} it is shown that the approximation
\begin{equation}\label{aprox2}
\langle\bar{\xi}_i(\mathbf{c})|\Lambda(\mathbf{c})g(\mathbf{c})\rangle
\approx-\lambda_i\langle\bar{\xi}_i(\mathbf{c})|g(\mathbf{c})\rangle, 
\quad i=2,3,4. 
\end{equation}
is consistent  with the equation for the correlation function, Eq. 
(\ref{eqphi}). Taking the scalar product with 
$\{\bar{\xi}_i\}_{i=1}^3$ in Eq. (\ref{eqphi}) an identity is obtained. 
Therefore, in contrast with approximation (\ref{aprox1}), the approximation 
given by Eq. (\ref{aprox2}) is fully consistent, i.e. it is compatible with 
both linear hydrodynamics and the equation for the two-particle correlations. 

To summarize, we have identified four modes. The first one (with the null 
eigenvalue) is trivial because is the one associated to the total number of 
particles. The second eigenvalue, $\lambda_2=\gamma$, vanishes in the elastic 
limit and is the one associated with the slowest excitations (at least in the 
elastic limit). For this reason, the second mode, $\Xi_2$, will 
be referred to as the hydrodynamic mode, in the following. The last two modes 
(one is the complex conjugate 
of the other) decay faster and will be called kinetic modes. Let us note that, 
although we have extended the number of fields to describe the excitations of 
the system, the number of slow modes remains the same (i.e. we have not 
adopted a kind 
of extended hydrodynamics approach as it could seem at first sight). Of course, 
these results are consistent with the ones of section 
\ref{spectralprop}. We obtain the same eigenvalue and, although the associated 
eigenfunctions are different, both modes are equivalent in the proper subspace. 
The differences in the modes are not important at the level of 
macroscopic hydrodynamics, but they are crucial at the level of two-particle 
correlations and, therefore, to identify the correct fluctuating hydrodynamic 
equations \cite{bgm12}. 

Let us evaluate the fluctuations of the total energy using the approximation 
given by (\ref{aprox2}). This can be done by taking the scalar products with 
$\langle\bar{\xi}_i(\mathbf{c}_1)\bar{\xi}_j(\mathbf{c}_2)|$ in the Eq. 
(\ref{eqphi}) for $i,j=2,3,4$, 
but, in contrast to the previous cases, these fluctuations are coupled to the 
ones of the pressure tensor. Nevertheless, we will see that this coupling 
disappears in the elastic limit (here we will restrict ourselves to $d=2$). 
In effect, multiplying Eq. (\ref{eqphi})
with $\langle\bar{\xi}_2(\mathbf{c}_1)\bar{\xi}_2(\mathbf{c}_2)|$ it is 
obtained 
\begin{equation}\label{ec22}
2\gamma\langle\bar{\xi}_2(\mathbf{c}_1)\bar{\xi}_2(\mathbf{c}_2)|
\phi(\mathbf{c}_1,\mathbf{c}_2)\rangle=
-\langle\bar{\xi}_2(\mathbf{c}_1)\bar{\xi}_2(\mathbf{c}_2)|
\widetilde{T}_0(\mathbf{c}_1,\mathbf{c}_2)\chi(\mathbf{c}_1)\chi(\mathbf{c}_2)
\rangle, 
\end{equation}
where approximation (\ref{aprox2}) has been used. For $d=2$ and to leading 
order ($\epsilon^2$ order in this case), we have
\begin{eqnarray}\label{i1}
\gamma\langle\bar{\xi}_2(\mathbf{c}_1)\bar{\xi}_2(\mathbf{c}_2)|
\phi(\mathbf{c}_1,\mathbf{c}_2)\rangle\approx\sqrt{\frac{\pi}{2}}\epsilon^2
\langle(2+2c_1^2)(2+2c_2^2)|\phi(\mathbf{c}_1,\mathbf{c}_2)\rangle\nonumber\\
=\sqrt{\frac{\pi}{2}}\epsilon^2
\left[4\langle c_1^2c_2^2|\phi(\mathbf{c}_1,\mathbf{c}_2)\rangle-12\right].  
\end{eqnarray}
The relation (\ref{phiVchi}) has been used to evaluate
\begin{equation}
\int d\mathbf{c}_1\phi(\mathbf{c}_1,\mathbf{c}_2)=
\int d\mathbf{c}_1c_i^2\phi(\mathbf{c}_1,\mathbf{c}_2)=-1.
\end{equation}
The right hand side of Eq. (\ref{ec22}) is evaluated in Appendix 
\ref{apendiceC} using the $\epsilon$ expansion of the Jenkins and Richman 
approximation, obtaining
\begin{equation}\label{i2}
\langle\bar{\xi}_2(\mathbf{c}_1)\bar{\xi}_2(\mathbf{c}_2)|
\widetilde{T}_0(\mathbf{c}_1,\mathbf{c}_2)\chi(\mathbf{c}_1)\chi(\mathbf{c}_2)\rangle
\approx-16\sqrt{2\pi}\epsilon^2. 
\end{equation}
By introducing Eqs. (\ref{i1}) and (\ref{i2}) into Eq. (\ref{ec22}), we have
\begin{equation}
\lim_{\alpha\to 1}\langle c_1^2c_2^2|\phi(\mathbf{c}_1,\mathbf{c}_2)\rangle=-1. 
\end{equation}
Finally, taking into account Eq. (\ref{flucE}), we can calculate the elastic 
limit of 
\begin{equation}
N\frac{\langle\delta\mathcal{E}^2\rangle}{\langle \mathcal{E}\rangle^2}=
\left[\int d\mathbf{c}c^4\chi(\mathbf{c})
+\int d\mathbf{c}_1\int d\mathbf{c}_2c_1^2c_2^2\phi(\mathbf{c}_1,\mathbf{c}_2)
\right]\to 1, 
\end{equation}
consistently with the results of \cite{bgm12}. 

\section{Fluctuations of the relevant global quantities}
\label{section5}

The structure of the modes derived above implies a coupling between the 
fluctuations of the total energy and the fluctuations of the pressure tensor 
for finite $\epsilon$. In this section, all these cross correlations will be 
evaluated. The fluctuating total pressure tensor is defined as
\begin{equation}
\mathcal{P}_{ij}(t)\equiv m\int d\mathbf{r}\int d\mathbf{v}V_iV_jF_1(x,t), 
\end{equation}
and its deviation can be written in the form indicated in Eq. (\ref{dA}). 
The correlations between $\delta\mathcal{E}$ and $\delta\mathcal{P}_{ij}$ can 
be calculated with the aid of Eq. (\ref{a1a2}), obtaining
\begin{equation}\label{flucEPij}
\langle\delta\mathcal{E}(t)\delta\mathcal{P}_{ij}(t)\rangle=\frac{m^2}{2}Nv_s^4
\left[\int d\mathbf{c}c^2c_ic_j\chi(\mathbf{c})+
\int d\mathbf{c}_1\int d\mathbf{c}_2c_1^2c_{2i}c_{2j}
\phi(\mathbf{c}_1,\mathbf{c}_2)\right].
\end{equation}
Analogously, it is 
\begin{equation}\label{flucPijPnm}
\langle\delta\mathcal{P}_{ij}(t)\delta\mathcal{P}_{nm}(t)\rangle=m^2Nv_s^4
\left[\int d\mathbf{c}c_ic_jc_nc_m\chi(\mathbf{c})+
\int d\mathbf{c}_1\int d\mathbf{c}_2c_{1i}c_{1j}c_{2n}c_{2m}
\phi(\mathbf{c}_1,\mathbf{c}_2)\right].
\end{equation}
This expression involves the first velocity moments of the correlation 
function, $\phi$. It is convenient to introduce the following notation
\begin{equation}
\mathbf{b}(\mathbf{c})=
\left[\begin{array}{c} 
1 \\ c^2 \\ c_xc_y \\ c_y^2
\end{array}\right], 
\end{equation}
allowing to express the moments in the following matrix form
\begin{equation}\label{defCij}
C_{ij}=\int d\mathbf{c}_1\int d\mathbf{c}_2b_i(\mathbf{c}_1)b_j(\mathbf{c}_2)
\phi(\mathbf{c}_1,\mathbf{c}_2),  
\end{equation}
that is trivially symmetric, i.e. $C_{ij}=C_{ji}$. 
%
In fact, the moments $\{C_{1j}\}_{j=1}^4$ can be easily calculated due to the 
conservation of the total number of particles. Taking into account Eq. 
(\ref{phiVchi}), we get 
\begin{equation}
C_{11}=-1, \quad C_{12}=-1, \quad C_{13}=-\frac{1}{2}\tilde{P}_{xy,s}, \quad 
C_{14}=-\frac{1}{2}\tilde{P}_{yy,s}. 
\end{equation}
To calculate the other $C_{ij}$ the scalar product 
$\langle\bar{\xi}_i(\mathbf{c}_1)\bar{\xi}_j(\mathbf{c}_2)|$ is taken in Eq. 
(\ref{eqphi}) and the approximation (\ref{aprox2}) introduced, obtaining
\begin{equation}\label{ec:124}
(\lambda_i+\lambda_j)\langle\bar{\xi}_i(\mathbf{c}_1)\bar{\xi}_j(\mathbf{c}_2)|
\phi(\mathbf{c}_1,\mathbf{c}_2)\rangle=
\langle\bar{\xi}_i(\mathbf{c}_1)\bar{\xi}_j(\mathbf{c}_2)|
\widetilde{T}_0(\mathbf{c}_1,\mathbf{c}_2)
\chi(\mathbf{c}_1)\chi(\mathbf{c}_2)\rangle, \quad i,j=2,3,4. 
\end{equation}
Actually, there are only $6$ independent equations, because of the relation 
between the third and fourth modes. As the scalar products 
$\langle\bar{\xi}_i(\mathbf{c}_1)\bar{\xi}_j(\mathbf{c}_2)|
\phi(\mathbf{c}_1,\mathbf{c}_2)\rangle$ (see Eq. (\ref{defxi}) and 
(\ref{defCij})) can be written in terms of the $C_{ij}$ coefficients through
\begin{equation}\label{ecxixi}
\langle\bar{\xi}_i(\mathbf{c}_1)\bar{\xi}_j(\mathbf{c}_2)|
\phi(\mathbf{c}_1,\mathbf{c}_2)\rangle
=\sum_{l=1}^4\xi_{il}\xi_{jl}C_{ll}
+\sum_{k>l=1}^4(\xi_{ik}\xi_{jl}+\xi_{il}\xi_{jk})C_{kl}, 
\end{equation}
Eq. (\ref{ec:124}) define a linear system of six 
equations for the six unknown coefficients 
$\{C_{22},C_{23},C_{24},C_{33},C_{34},C_{44}\}$ 
(remember that $\{C_{1j}\}_{j=1}^4$ are known).

The calculation leading to the expressions of the coefficients $C_{ij}$ are 
detailed in Appendix \ref{apendiceC}. Since the 
$\langle\bar{\xi}_i(\mathbf{c}_1)\bar{\xi}_j(\mathbf{c}_2)|
\widetilde{T}_0(\mathbf{c}_1,\mathbf{c}_2)
\chi(\mathbf{c}_1)\chi(\mathbf{c}_2)\rangle$ are evaluated using 
the Jenkins and Richman distribution function for $d=2$, in the following all 
the results are restricted to this dimension. To order $\epsilon^2$ the 
obtained expressions are 
\begin{eqnarray}\label{cij}
C_{22}=-1+\frac{27}{32}\epsilon^2, \quad C_{23}=\frac{5}{8\sqrt{2}}\epsilon, 
\quad C_{24}=-\frac{1}{2}+\frac{51}{64}\epsilon^2, \nonumber\\
\quad C_{33}=-\frac{23}{256}\epsilon^2, 
\quad C_{34}=\frac{5}{16\sqrt{2}}\epsilon, 
\quad C_{44}=-\frac{1}{4}+\frac{145}{256}\epsilon^2. 
\end{eqnarray}
The one-particle averages that appear in the equations of the fluctuations, 
$\langle b_i(\mathbf{c})b_j(\mathbf{c})|\chi(\mathbf{c})\rangle$, can be 
calculated in the same approximation, obtaining
\begin{eqnarray}\label{momentos}
\langle c^4|\chi(\mathbf{c})\rangle=2+\frac{1}{2}\epsilon^2, 
\quad\langle c^2c_xc_y|\chi(\mathbf{c})\rangle=-\frac{3}{2\sqrt{2}}\epsilon, 
\quad\langle c^2c_y^2|\chi(\mathbf{c})\rangle=1-\frac{1}{2}\epsilon^2, 
\nonumber\\
\langle c_x^2c_y^2|\chi(\mathbf{c})\rangle=\frac{1}{4}(1+\epsilon^2), \quad
\langle c_xc_y^3|\chi(\mathbf{c})\rangle=-\frac{3}{4\sqrt{2}}\epsilon,  \quad
\langle c_y^4|\chi(\mathbf{c})\rangle=\frac{3}{4}(1-\epsilon^2). 
\end{eqnarray}

To express the final result in a compact notation it is useful to introduce 
the matrix elements
\begin{equation}\label{defB}
B_{ij}(0)\equiv
\int d\mathbf{c}_1\int d\mathbf{c}_2b_i(\mathbf{c}_1)b_j(\mathbf{c}_2)
\psi(\mathbf{c}_1,\mathbf{c}_2,0)
=\langle b_i(\mathbf{c})b_j(\mathbf{c})|\chi(\mathbf{c})\rangle+C_{ij}.  
\end{equation}
By substituting Eqs. (\ref{cij}) and (\ref{momentos}) into the equation above, 
the expansion to second order in $\epsilon$ of $B(0)$ is obtained, 
\begin{equation}\label{predictionB}
B(0)=\left[\begin{array}{cccc}
0 & 0 & 0 & 0 \\
0 & 1+\frac{199\epsilon^2}{64} 
& -\frac{7\epsilon}{8\sqrt{2}} & \frac{1}{2}+\frac{19\epsilon^2}{64} \\
0 & -\frac{7\epsilon}{8\sqrt{2}} 
& \frac{1}{4}+\frac{41\epsilon^2}{256} & -\frac{7\epsilon}{16\sqrt{2}} \\
0 & \frac{1}{2}+\frac{19\epsilon^2}{64} 
& -\frac{7\epsilon}{16\sqrt{2}} & \frac{1}{2}-\frac{47\epsilon^2}{256}
\end{array}\right]. 
\end{equation}
Finally, taking into account Eqs. (\ref{defB}), (\ref{flucE}), 
(\ref{flucEPij}), and (\ref{flucPijPnm}), we can express all the correlation 
functions in terms of $B_{ij}(0)$, 
\begin{eqnarray}
\langle\delta\mathcal{E}^2(t)\rangle=\frac{m^2}{4}Nv_s^4B_{22}(0), \quad 
\langle\delta\mathcal{E}(t)\delta\mathcal{P}_{xy}(t)\rangle
=\frac{m^2}{2}Nv_s^4B_{23}(0), \nonumber\\
\langle\delta\mathcal{E}(t)\delta\mathcal{P}_{yy}(t)\rangle
=\frac{m^2}{2}Nv_s^4B_{24}(0), \quad
\langle\delta\mathcal{P}_{xy}^2(t)\rangle=m^2Nv_s^4B_{33}(0),\nonumber\\
\langle\delta\mathcal{P}_{xy}(t)\delta\mathcal{P}_{yy}(t)\rangle
=m^2Nv_s^4B_{34}(0), \quad
\langle\delta\mathcal{P}_{yy}^2(t)\rangle
=m^2Nv_s^4B_{44}(0). 
\end{eqnarray}
It is worth to remark 
that, although the system has been solved consistently to $\epsilon^2$ order, 
the expressions for the correlation functions are not the exact power expansion 
of the correlation functions. This is so because the Jenkins and Richman 
approximation to $\epsilon^2$ order is not the exact expansion of the 
distribution \cite{sgn96}. 


Finally, let us calculated the two-time correlation functions between the 
already considered global quantities. Using Eq. (\ref{a1a2t}) we arrive to the 
generalization of Eqs. (\ref{flucE}), (\ref{flucEPij}) and (\ref{flucPijPnm}) 
for two times
\begin{eqnarray}\label{flucEt}
\langle\delta\mathcal{E}(t)\delta\mathcal{E}(t')\rangle=\frac{m^2}{4}Nv_s^4
\int d\mathbf{c}_1\int d\mathbf{c}_2c_1^2c_2^2
\psi(\mathbf{c}_1,\mathbf{c}_2;s-s'), \\
\label{flucEPijt}
\langle\delta\mathcal{E}(t)\delta\mathcal{P}_{ij}(t')\rangle=\frac{m^2}{2}Nv_s^4
\int d\mathbf{c}_1\int d\mathbf{c}_2c_1^2c_{2i}c_{2j}
\psi(\mathbf{c}_1,\mathbf{c}_2,s-s'), \\
\label{flucPijEt}
\langle\delta\mathcal{P}_{ij}(t)\delta\mathcal{E}(t')\rangle=\frac{m^2}{2}Nv_s^4
\int d\mathbf{c}_1\int d\mathbf{c}_2c_{1i}c_{1j}c_2^2
\psi(\mathbf{c}_1,\mathbf{c}_2,s-s'), \\
\label{flucPijPnmt}
\langle\delta\mathcal{P}_{ij}(t)\delta\mathcal{P}_{nm}(t')\rangle=m^2Nv_s^4
\int d\mathbf{c}_1\int d\mathbf{c}_2c_{1i}c_{1j}c_{2n}c_{2m}
\psi(\mathbf{c}_1,\mathbf{c}_2,s-s'). 
\end{eqnarray}
Again, it is convenient to define the matrix elements
\begin{equation}
B_{ij}(s)\equiv\int d\mathbf{c}_1\int d\mathbf{c}_2
b_i(\mathbf{c}_1)b_j(\mathbf{c}_2)
\psi(\mathbf{c}_1,\mathbf{c}_2,s).  
\end{equation}
Inserting the formal expression of $\psi(s)$ 
\begin{equation}
\psi(\mathbf{c}_1,\mathbf{c}_2,s)=e^{s\Lambda(\mathbf{c}_1)}
[\chi(\mathbf{c}_1)\delta(\mathbf{c}_{12})+\phi(\mathbf{c}_1,\mathbf{c}_2)], 
\end{equation}
into the above equations, and taking into account that the functions 
$\{b_i(\mathbf{c})\}_{i=1}^4$ can be written in terms of the functions 
$\{\bar{\xi}_i(\mathbf{c})\}_{i=1}^4$, the correlation functions can be 
evaluated explicitly by using the approximation (\ref{aprox2}), with the 
result  
\begin{equation}\label{aiajt}
B_{ij}(s)=\sum_{l=1}^4\sum_{l=1}^4Q^{-1}_{ik}Q_{kl}e^{\lambda_ks}B_{lj}(0), \quad s>0.  
\end{equation}
Here we have introduced the matrix $Q$
\begin{equation}
\left[\begin{array}{c} \bar{\xi}_1(\mathbf{c})\\ \bar{\xi}_2(\mathbf{c})\\
\bar{\xi}_3(\mathbf{c})\\ \bar{\xi}_4(\mathbf{c})\\ 
\end{array}\right]=Q\mathbf{b}(\mathbf{c}), 
\end{equation}
and its inverse, $Q^{-1}$, that can be identified through Eq. (\ref{defxi}). 
To order $\epsilon^2$ we have
\begin{equation}
Q=
\left[\begin{array}{cccc} 
\xi_{11} & \xi_{12} & \xi_{13} & \xi_{14} \\
\xi_{21} & \xi_{22} & \xi_{23} & \xi_{24} \\
\xi_{31} & \xi_{32} & \xi_{33} & \xi_{34} \\
\xi_{41} & \xi_{42} & \xi_{43} & \xi_{44} 
\end{array}\right]\approx
\left[\begin{array}{cccc}
1 & 0 & 0 & 0 \\
2-\frac{3\epsilon^2}{4} & 2-\frac{11\epsilon^2}{4} 
& -2\sqrt{2}\epsilon & 2\epsilon^2 \\
-\frac{9\epsilon^2}{16} 
& -1+\frac{\epsilon^2}{16}-\imath\frac{\epsilon}{2\sqrt{2}} 
& \frac{\epsilon}{2\sqrt{2}}-2\imath\left(1-\frac{\epsilon^2}{64}\right) & 2 \\
-\frac{9\epsilon^2}{16} 
& -1+\frac{\epsilon^2}{16}+\imath\frac{\epsilon}{2\sqrt{2}} 
& \frac{\epsilon}{2\sqrt{2}}+2\imath\left(1-\frac{\epsilon^2}{64}\right) & 2 
\end{array}\right]. 
\end{equation}
Obviously, the correlation functions given by 
(\ref{aiajt}) fulfill the initial conditions. Moreover, they are a linear 
combination of the two modes $\lambda_2$ and $\lambda_3$ 
($\lambda_4=\lambda_3^*$). The complete expressions for all the correlation 
functions are very lengthy and here we only write explicitly the expressions 
for the two-time autocorrelation function of the energy and pressure tensor, 
\begin{equation}
\langle\delta\mathcal{E}(t)\delta\mathcal{E}(0)\rangle=
\frac{m^2}{4}Nv_s^4B_{22}(s), \quad
\langle\delta\mathcal{P}_{xy}(t)\delta\mathcal{P}_{xy}(0)\rangle
=m^2Nv_s^4B_{33}(s),
\end{equation} 
with $B_{22}(s)$ and $B_{33}(s)$ given by Eq. (\ref{aiajt}), i.e. 
\begin{equation}\label{b22s}
B_{22}(s)
=\left(1+\frac{225\epsilon^2}{128}\right)e^{-\lambda_2 s}
-\frac{5\epsilon^2}{8}\cos{(\lambda_3^Is)}e^{-\lambda_3^R s}, 
\end{equation}
\begin{equation}\label{b33s}
B_{33}(s)=
\left[\left(\frac{1}{4}-\frac{3\epsilon^2}{256}\right)\cos{(\lambda_3^I s)}
+\frac{\sqrt{2}\epsilon}{32}\sin{(\lambda_3^I s)}\right]e^{-\lambda_3^R s}
+\frac{11\epsilon^2}{64}e^{-\lambda_2 s}. 
\end{equation}
We see that both functions have a hydrodynamic and a kinetic part. 
Nevertheless, the main contribution of $B_{22}(s)$ is the hydrodynamic one (the 
kinetic part is of order $\epsilon^2$), while the opposite occurs with 
$B_{33}(s)$.

\section{Simulation results}\label{section6}

We have performed Molecular Dynamics (MD) simulations of a two dimensional 
system of $N=2000$ inelastic hard disks of mass $m$ and diameter $\sigma$, in a 
square box of side $L$, corresponding to a number density $n_s=0.02\sigma^{-2}$. 
To generate the stationary USF state, Lees-Edwards boundary conditions 
\cite{le72} in the $y$-direction and periodic boundary conditions in the 
$x$-direction have been used. Once the steady state is reached, we have 
measured all the quantities studied in the previous section. The reported 
values have been averaged over 300 trajectories, and also on time, over a 
period of about 150 collisions per particle. This has been done for different 
values of the inelasticity, $\alpha$. The shear rate was in all cases 
$a=6.32\times10^{-3}(T(0)/m)^{1/2}\sigma^{-1}$, where $T(0)$ is the initial 
temperature.

\begin{figure}
\includegraphics[angle=0,width=0.9\linewidth]
{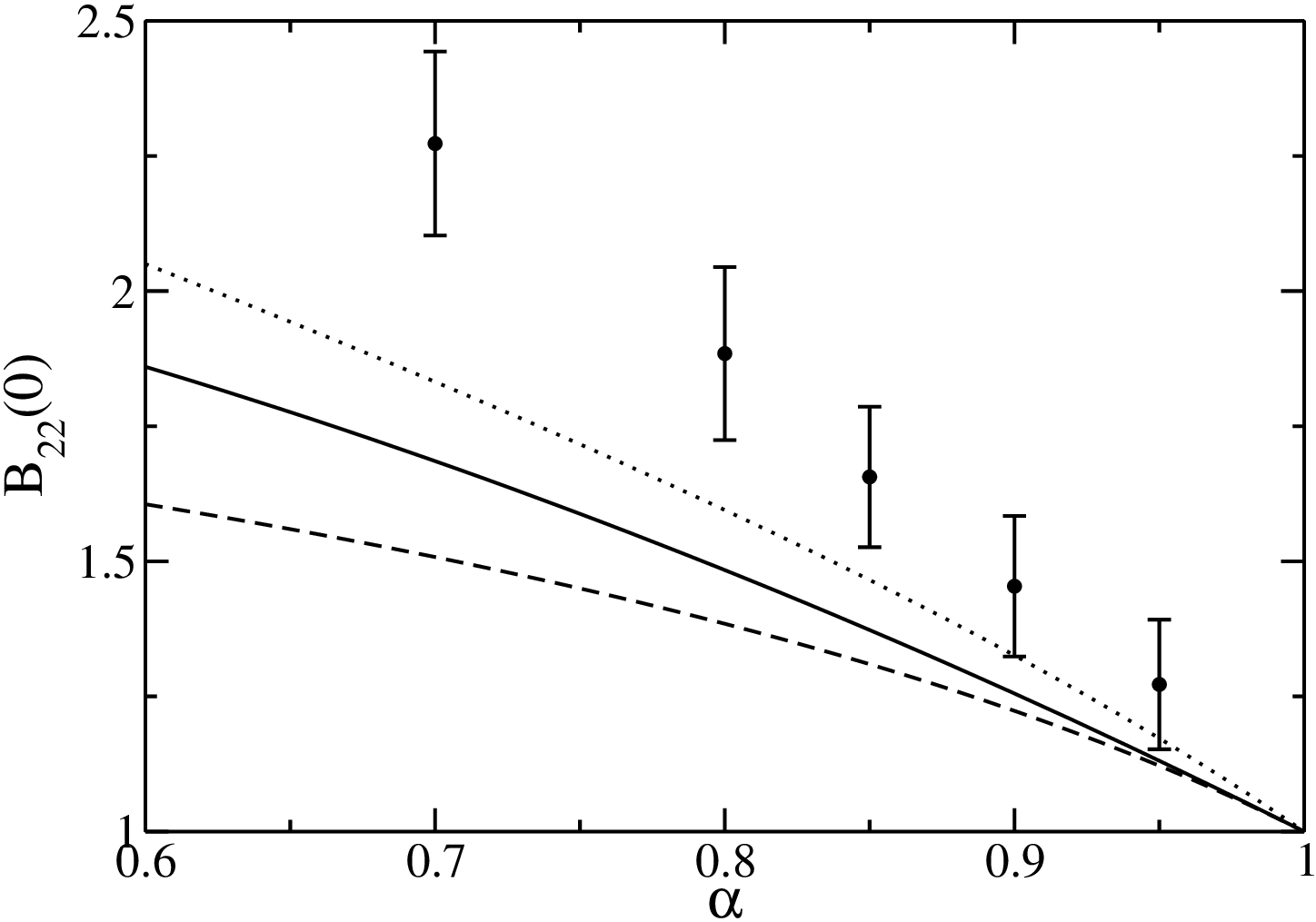}
\caption{Dimensionless matrix element $B_{22}(0)$ as  function of the 
restitution coefficient, $\alpha$, for a system of $N=2000$ hard disks. The 
symbols (dots) are the simulation results, the solid line is the theoretical 
prediction given by Eq. (\ref{predictionB}), the dashed line is the prediction 
using fluctuating hydrodynamics, and the dotted line the improved prediction 
given in \cite{bgm12}. }
\label{fig1}
\end{figure}

\begin{figure}
\begin{minipage}{0.48\linewidth}
\begin{center}
\includegraphics[angle=0,width=0.95\linewidth]{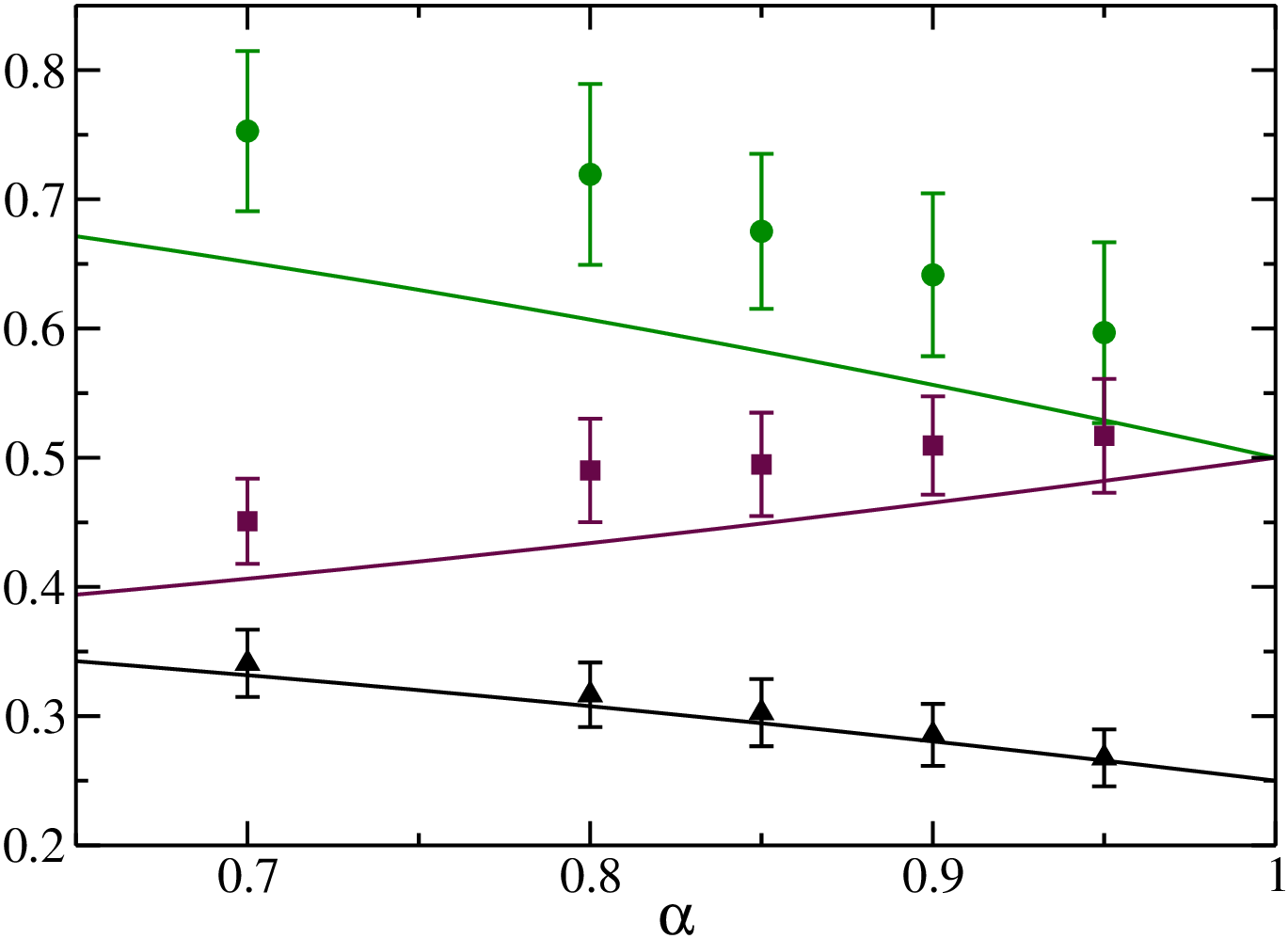}
\end{center}
\end{minipage}
\begin{minipage}{0.48\linewidth}
\begin{center}
\includegraphics[angle=0,width=0.95\linewidth]{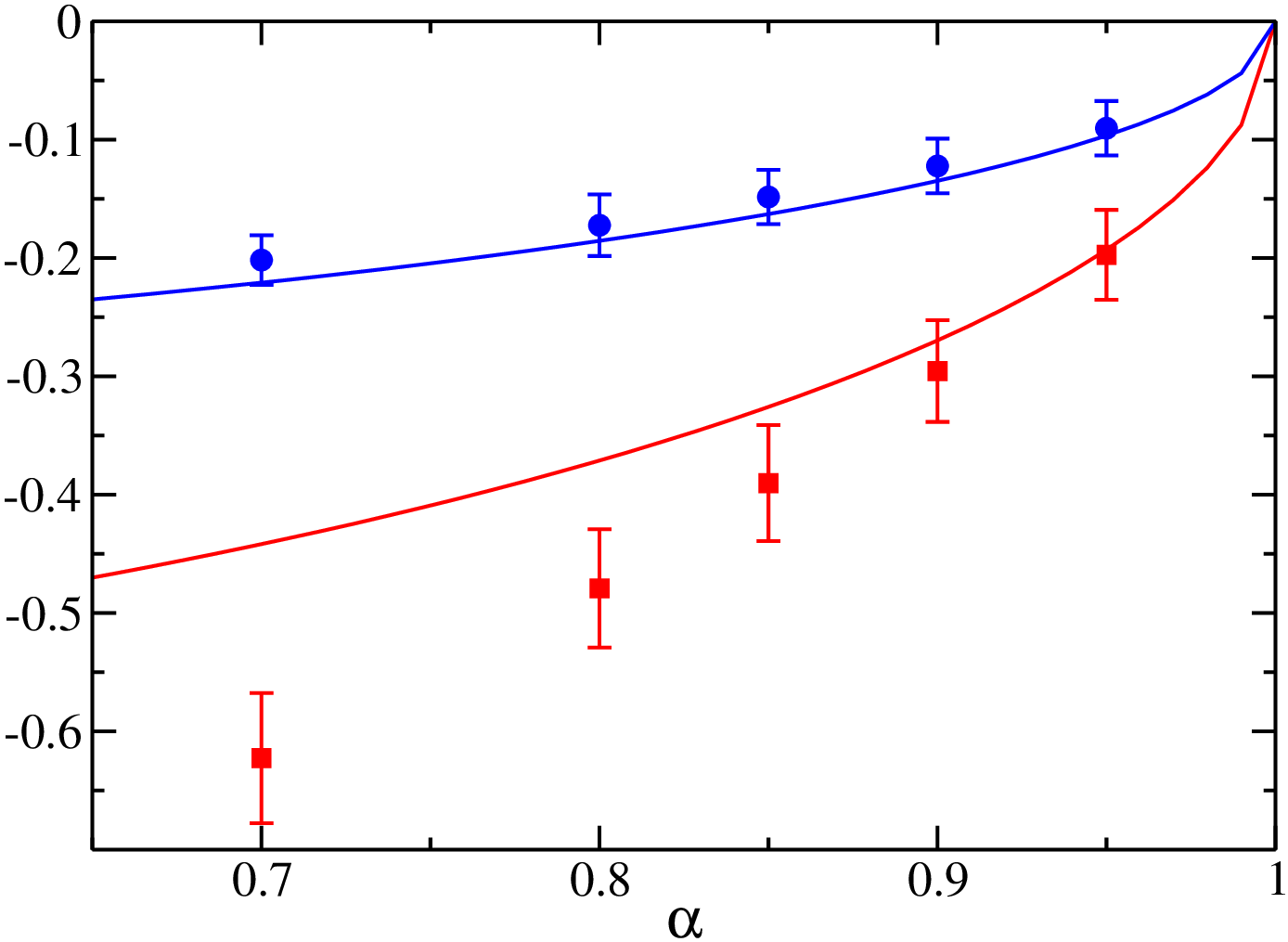}
\end{center}
\end{minipage}
\caption{(Color on line) Dimensionless matrix elements of $B(0)$. The symbols 
are the simulation data and the solid lines correspond to the expansion to 
second order in $\epsilon$ given in Eq. (\ref{predictionB}). In the left 
figure, the circles, squares and triangles correspond to $B_{24}(0)$, 
$B_{44}(0)$ and $B_{33}(0)$ respectively. In the right figure, the circles and 
squares correspond to $B_{34}(0)$ and $B_{23}(0)$ respectively.  }\label{fig2}
\end{figure}

In Fig. \ref{fig1} we plot the quantity $B_{22}(0)$ as a function of the 
inelasticity. The symbols are the simulation results and the solid line the 
theoretical prediction given by Eq. (\ref{predictionB}). We have also plotted 
the theoretical prediction of the fluctuating hydrodynamic approach (dashed 
line) and the improved one (dotted-line) taking into account rheological 
effects in the 
viscosity \cite{bgm12}. As it is shown in the figure, the last one is very 
close to the prediction given by (\ref{predictionB}). In Fig. \ref{fig2} we 
plot the rest of matrix elements of $B$ as a function of $\alpha$. The solid 
lines are the theoretical predictions and the symbols are the simulation 
results. While the agreement is very good for $B_{23}(0)$, $B_{33}(0)$ and 
$B_{34}(0)$, we find some discrepancies for $B_{22}(0)$, $B_{24}(0)$ and 
$B_{44}(0)$ as the inelasticity increases. 

\begin{figure}
\begin{minipage}{0.48\linewidth}
\begin{center}
\includegraphics[angle=0,width=0.95\linewidth]{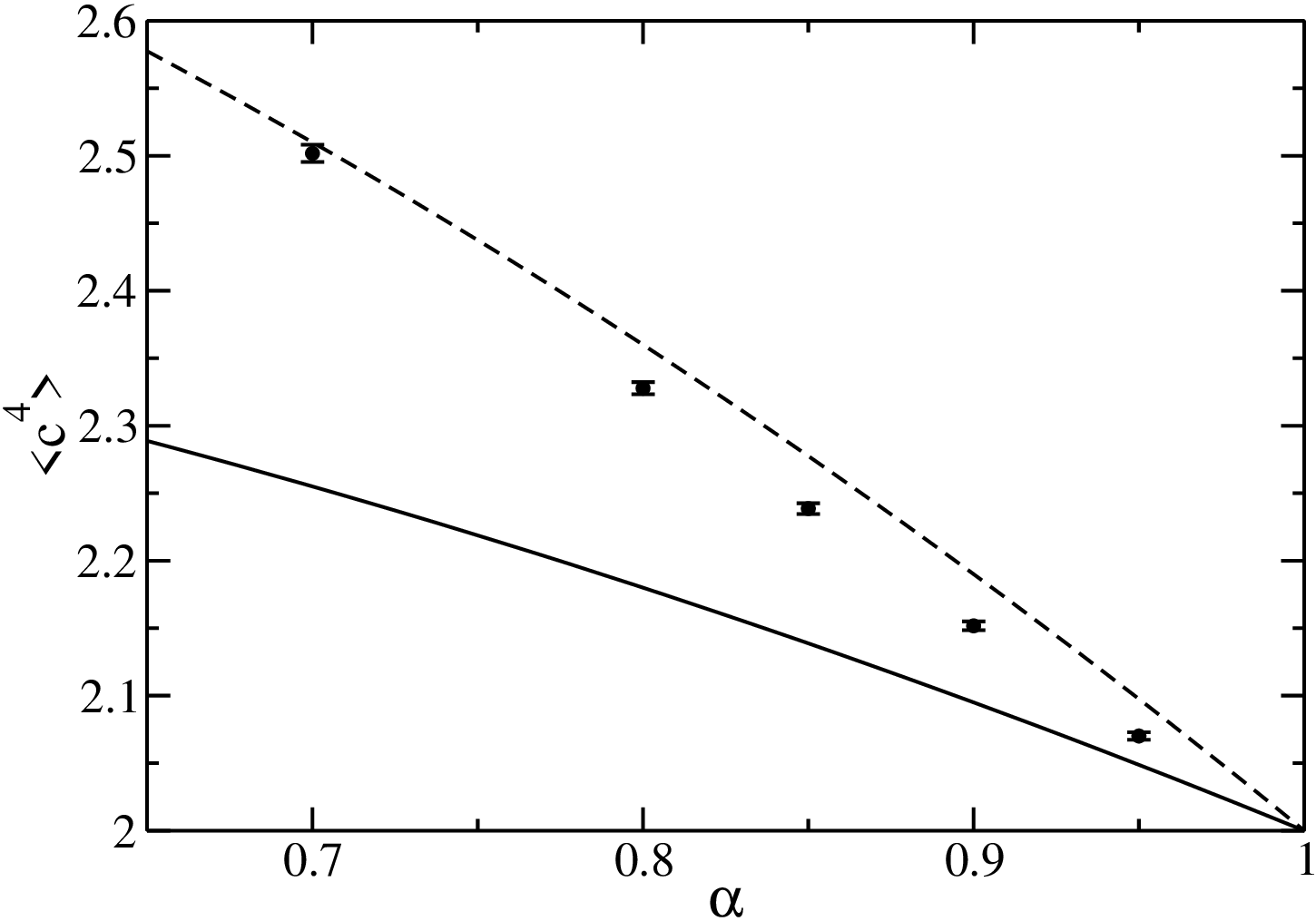}
\end{center}
\end{minipage}
\begin{minipage}{0.48\linewidth}
\begin{center}
\includegraphics[angle=0,width=0.95\linewidth]{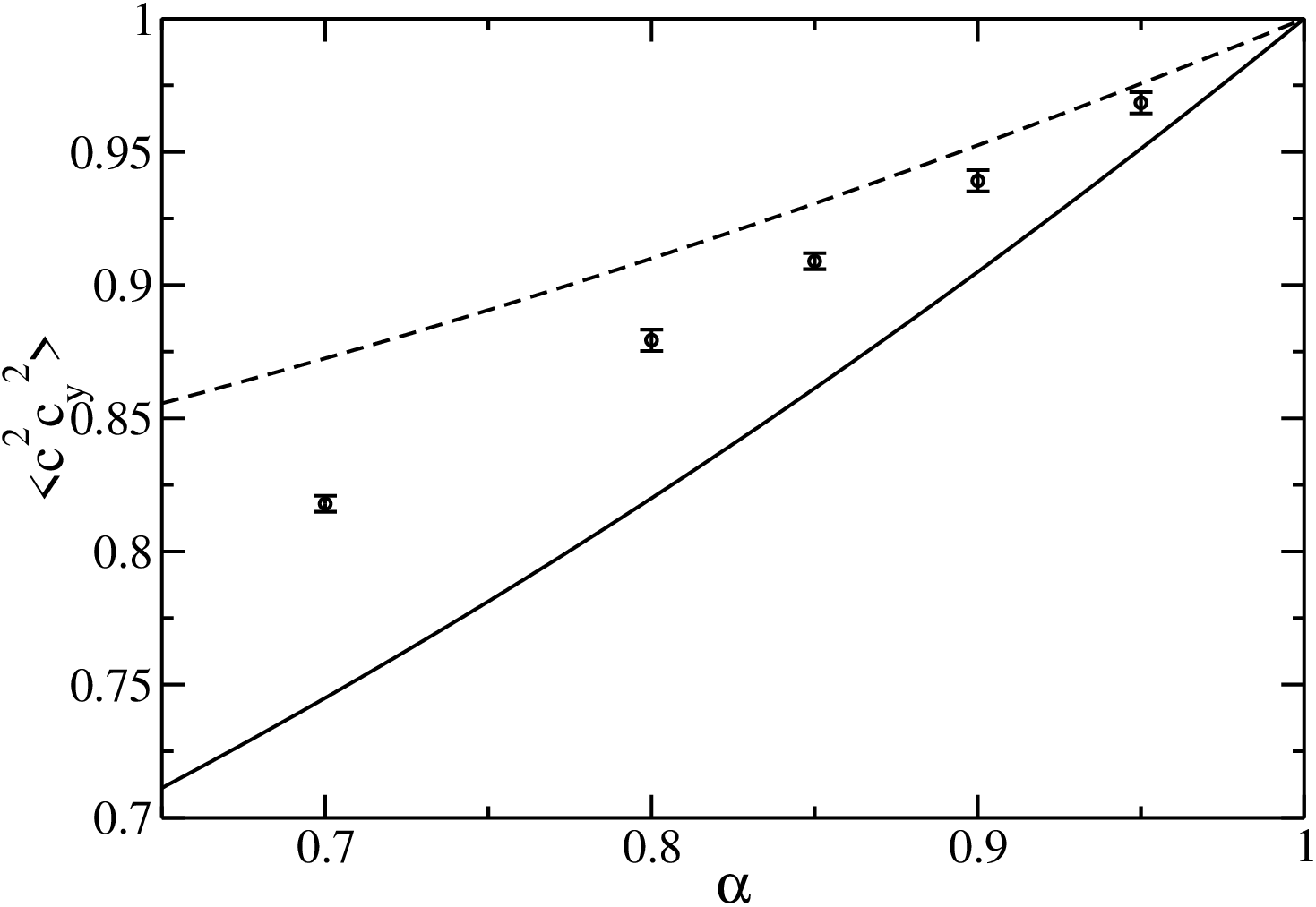}
\end{center}
\end{minipage}
\caption{One-particle averages $\langle c^4\rangle$ (left) and 
$\langle c^2c_y^2\rangle$ (right). The simulation data (symbols) are compared 
to the predictions of Jenkins and Richman \cite{jr88} (solid line) and to the 
BGK model \cite{brm97} (dashed line).}\label{fig3}
\end{figure}

\begin{figure}
\begin{minipage}{0.48\linewidth}
\begin{center}
\includegraphics[angle=0,width=0.95\linewidth]{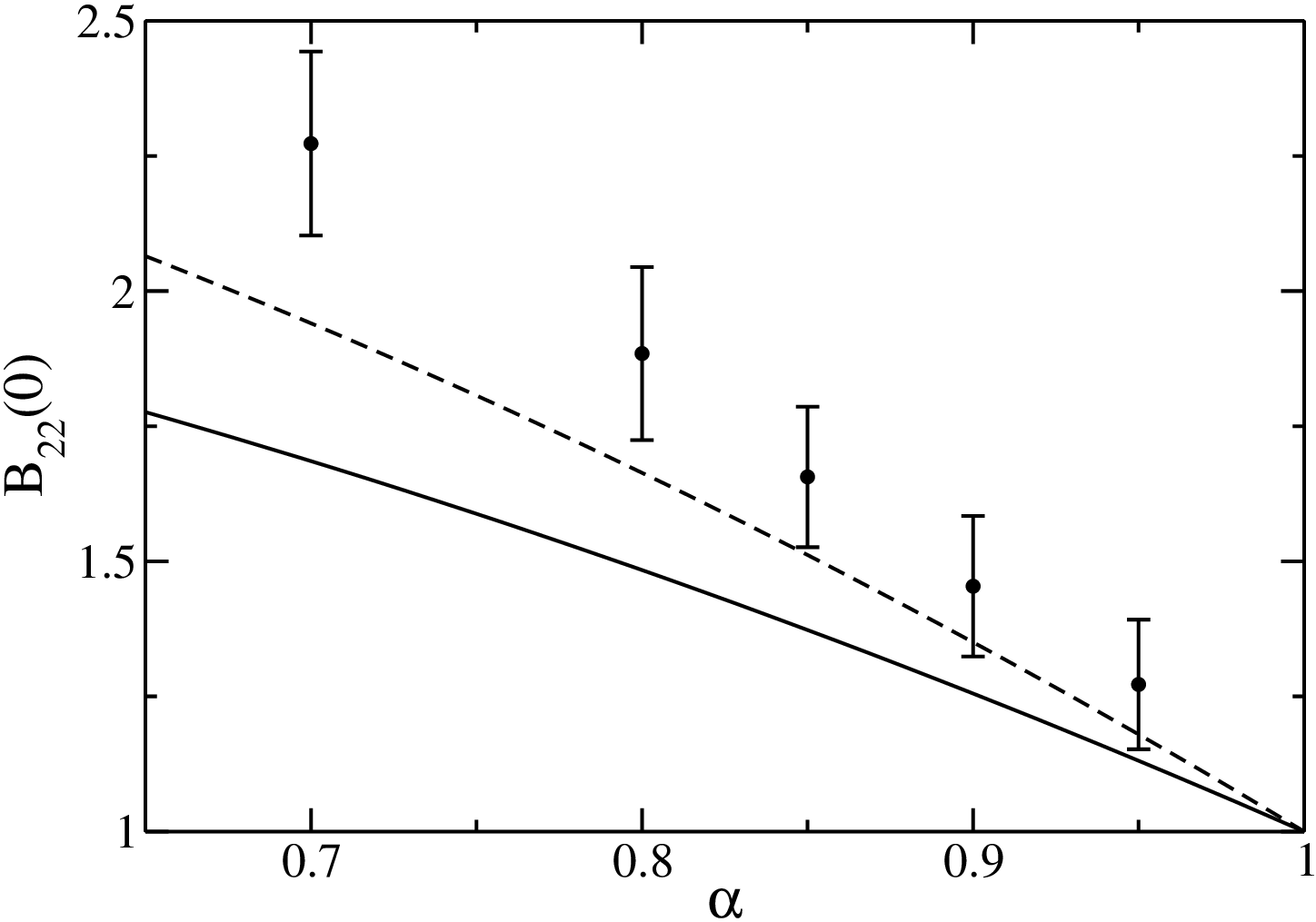}
\end{center}
\end{minipage}
\begin{minipage}{0.48\linewidth}
\begin{center}
\includegraphics[angle=0,width=0.95\linewidth]{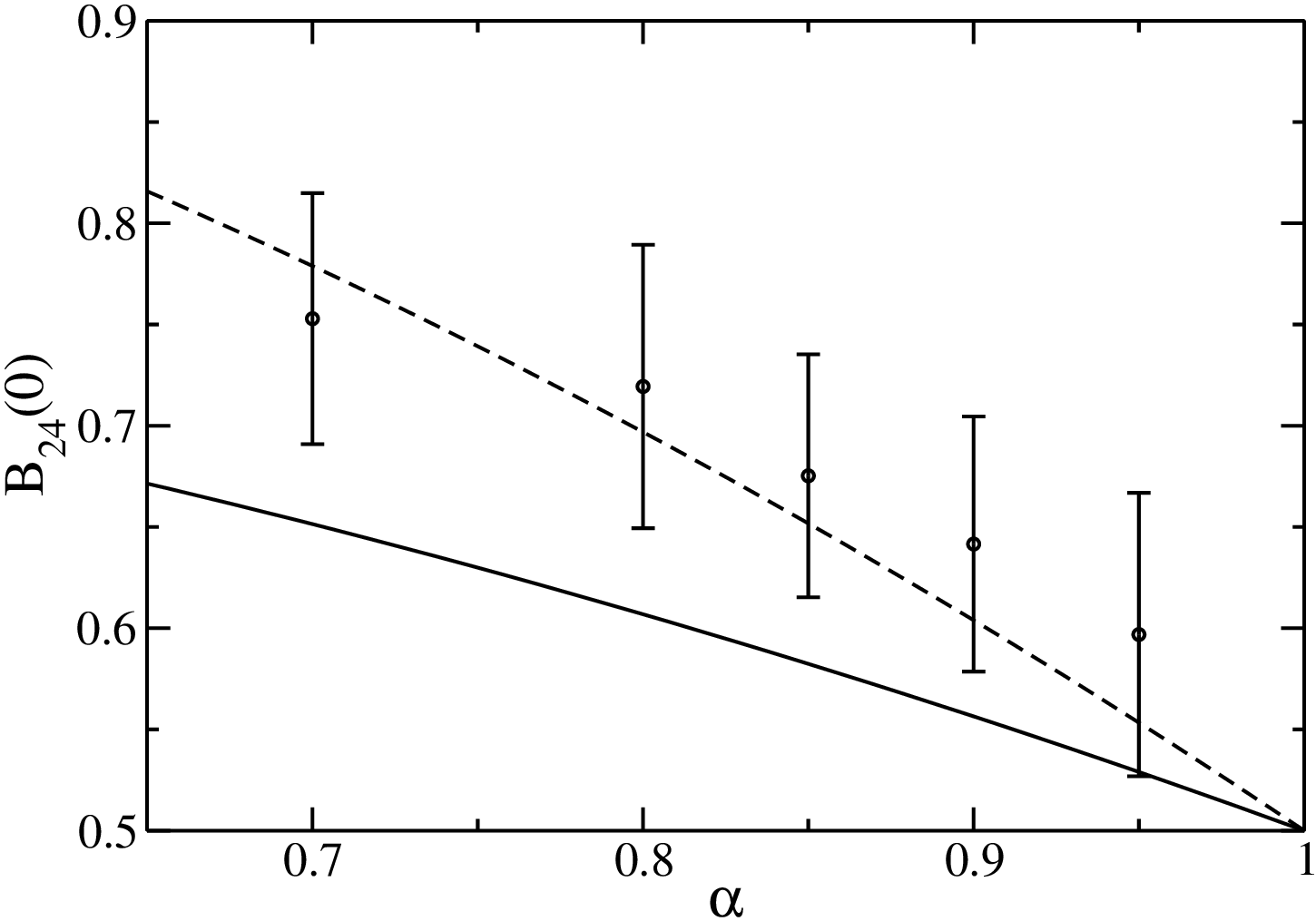}
\end{center}
\end{minipage}
\caption{Comparison of the simulation results for $B_{22}(0)$ and $B_{24}(0)$ 
with the theoretical predictions using the Jenkins and Richman approximation 
\cite{jr88} (solid lines) or the BGK model \cite{brm97} (dashed line) for the 
one-particle moments.}
\label{fig4}
\end{figure}

As the $B_{ij}(0)$ coefficients have two components, the one-particle component 
and the correlation function component, we have measured the one-particle 
moments implied in order to study the origin of the discrepancies. In Fig. 
\ref{fig3}, we plot $\langle c^4\rangle$ and $\langle c^2c_y^2\rangle$. As it 
is seen in the figure, there are important differences between the simulation 
results (points) and the Jenkins and Richman approximation (solid line). We 
also plot the theoretical prediction of the moments to $\epsilon^2$ order using 
the BGK model \cite{brm97} finding a remarkably better agreement. Its explicit 
expressions are
\begin{equation}
\langle c^4\rangle_{BGK}=2+\epsilon^2, \quad 
\langle c^2c_y^2\rangle_{BGK}=1-\frac{\epsilon^2}{4}. 
\end{equation}
The rest of moments are accurately described by the Jenkins and Richman 
approximation. In fact, they coincide with the BGK ones apart from 
$\langle c_x^2c_y^2\rangle$, for which the Jenkins and Richman approximation 
goes better than the BGK prediction. 
In Fig. \ref{fig4} we plot $B_{22}(0)$ and 
$B_{24}(0)$ using the one-particle moments of the BGK model, finding that this 
increases considerably the agreement with the simulation results. Hence, we 
can conclude that the agreement between the simulation results and 
the theoretical predictions (considering the most accurate expression for the 
one-particle moments) is excellent for all the coefficients for the range of 
inelasticities considered. The only exception is $B_{44}$ for which the 
agreement is moderately good. 

\begin{figure}
\begin{minipage}{0.48\linewidth}
\begin{center}
\includegraphics[angle=0,width=0.95\linewidth]
{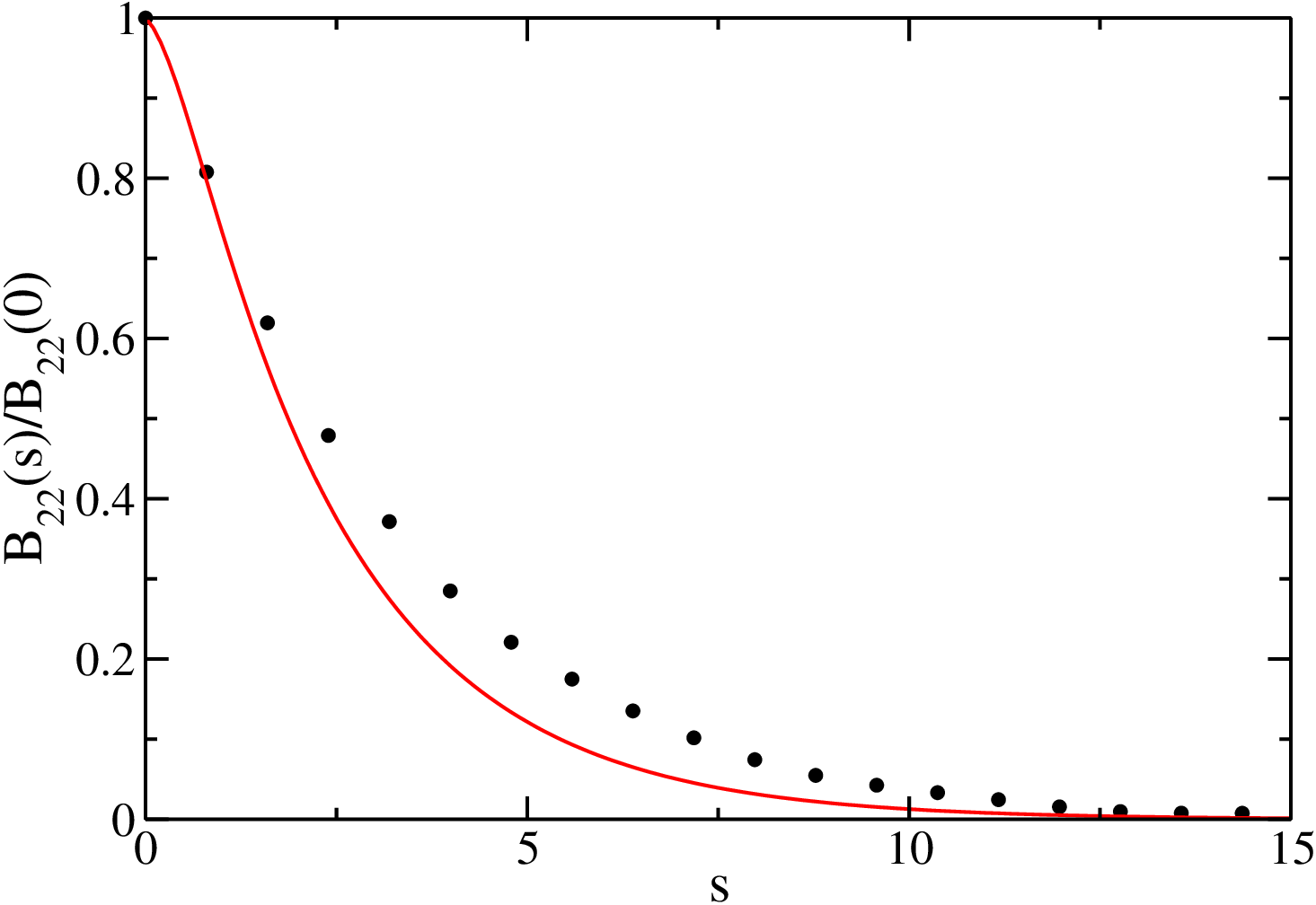}
\end{center}
\end{minipage}
\begin{minipage}{0.48\linewidth}
\begin{center}
\includegraphics[angle=0,width=0.95\linewidth]
{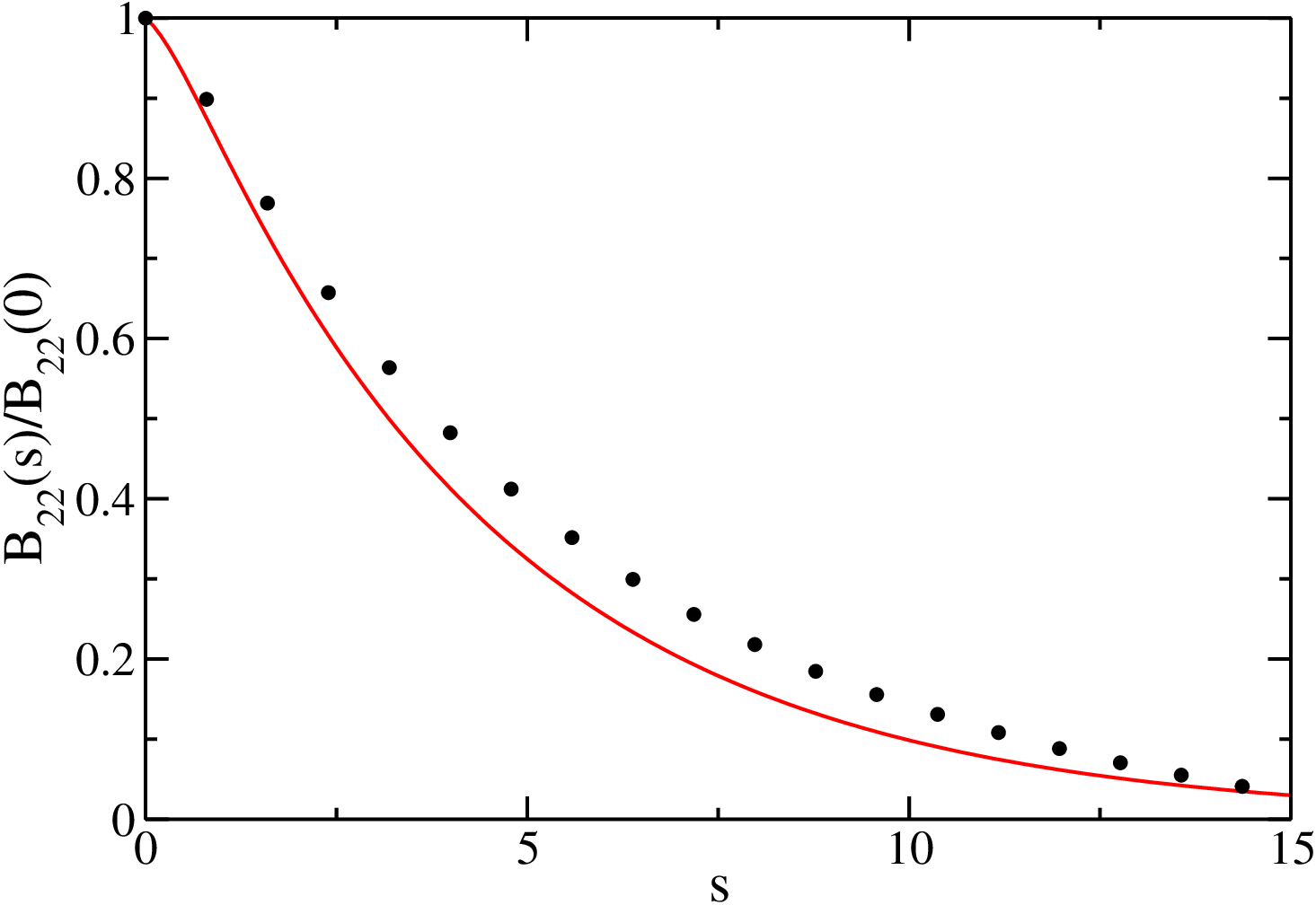}
\end{center}
\end{minipage}
\caption{(Color on line) Decay of $B_{22}(s)/B_{22}(0)$ for a system with 
$\alpha=0.80$ (left) and $\alpha=0.90$ (right). The solid lines (red) are the 
predictions given by Eq. (\ref{b22s}), and the symbols are the simulation 
results.}\label{fig5}
\end{figure}

\begin{figure}
\begin{minipage}{0.48\linewidth}
\begin{center}
\includegraphics[angle=0,width=0.95\linewidth]{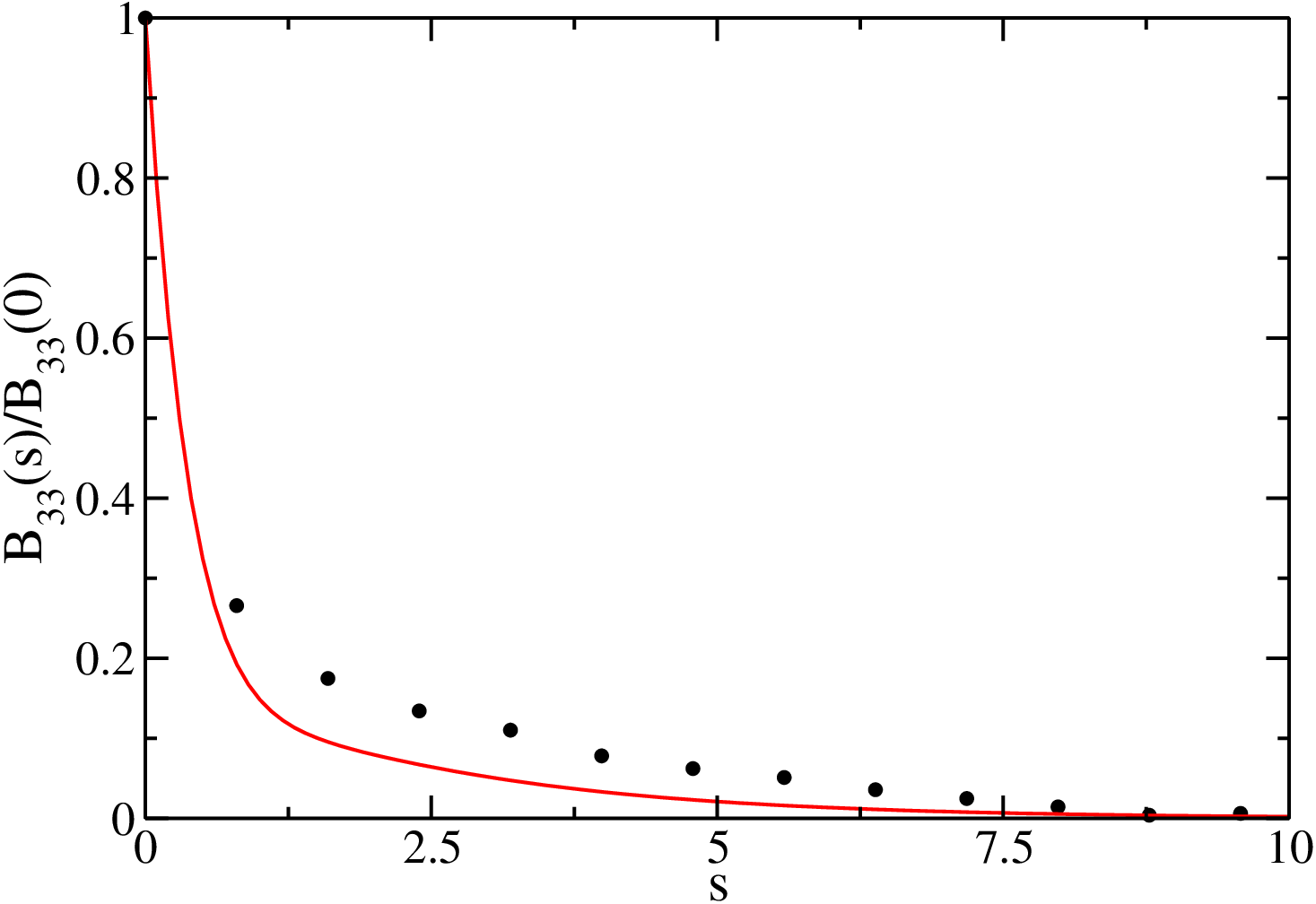}
\end{center}
\end{minipage}
\begin{minipage}{0.48\linewidth}
\begin{center}
\includegraphics[angle=0,width=0.95\linewidth]{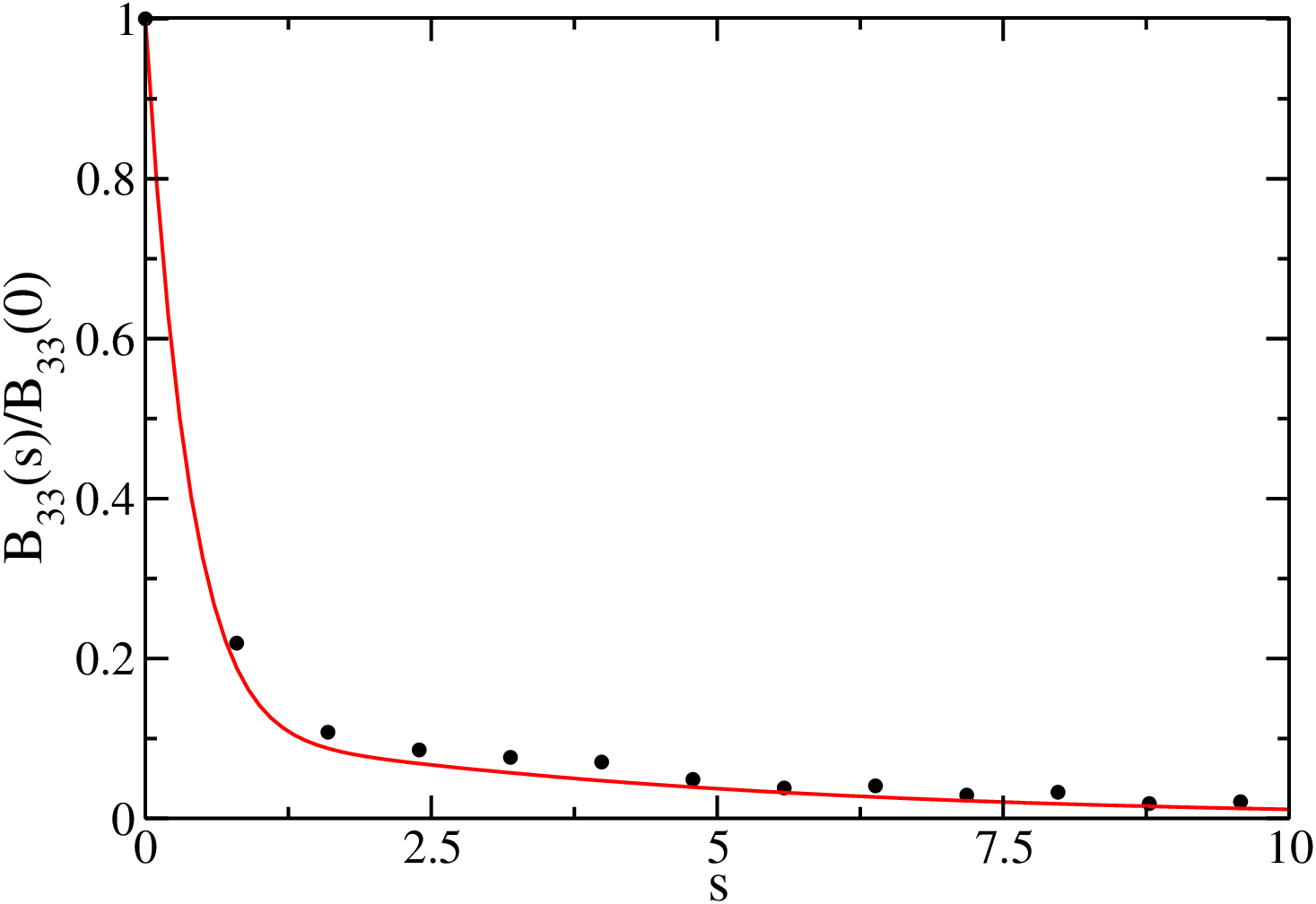}
\end{center}
\end{minipage}
\caption{(Color on line) Decay of $B_{33}(s)/B_{33}(0)$ for a system with 
$\alpha=0.80$ (left) and $\alpha=0.90$ (right). The solid lines (red) are the 
predictions given by Eq. (\ref{b33s}), and the symbols are the simulation 
results.}\label{fig6}
\end{figure}

\begin{figure}
\includegraphics[angle=0,width=0.9\linewidth]
{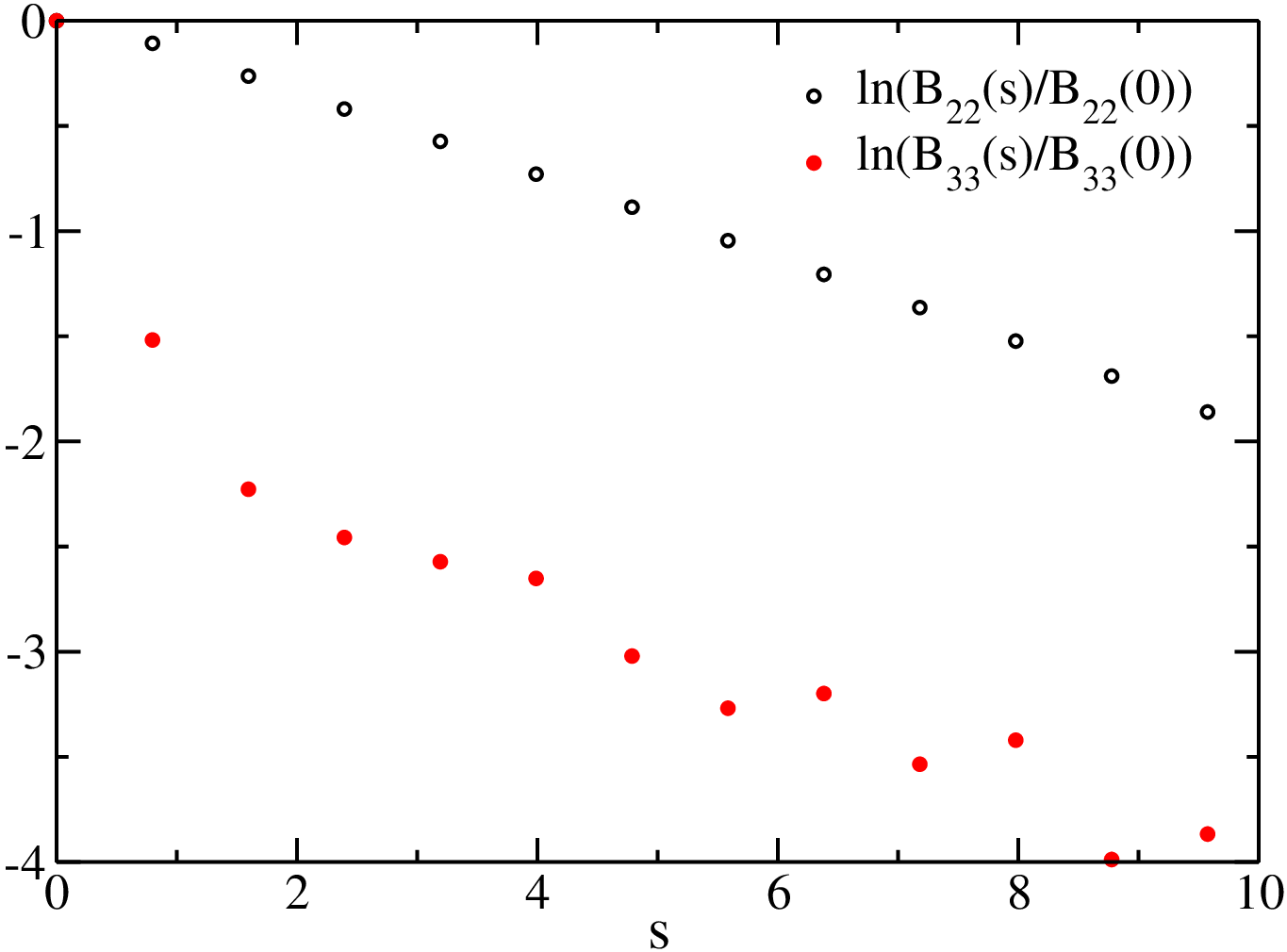}
\caption{Comparison of the decay in the long time limit of the correlations 
$B_{22}(s)/B_{22}(0)$ and $B_{33}(s)/B_{33}(0)$ for a system with $\alpha=0.90$.}
\label{fig7}
\end{figure}

Finally, we have also measured the two-time correlation functions. Fig. 
\ref{fig5} shows the evolution of $B_{22}(s)/B_{22}(0)$, for systems with 
$\alpha=0.80$ (left) and $\alpha=0.90$ (right). In Fig. \ref{fig6} the decay 
of  $B_{33}(s)/B_{33}(0)$ has been plotted for the same values of the 
inelasticity. As the correlation functions are a combination of two 
exponentials, it is difficult to make a detail comparison between the 
theoretical prediction and the simulation results. Nevertheless, it is observed 
that $B_{33}(s)$ decays faster than $B_{22}(s)$, as predicted by Eqs. 
(\ref{b22s}) and (\ref{b33s}). $B_{33}(s)$ has a hydrodynamic part that is of 
order $\epsilon^2$, while the kinetic part is of order unity (the contrary 
occurs for $B_{22}(s)$). We have also seen that, in the long-time limit, both 
correlation functions decay in the same way (with the hydrodynamic mode). In 
order to see the long time behavior of the functions, they have been plotted 
in a logarithmic scale. This is done in Fig. \ref{fig7} for a system with 
$\alpha=0.90$, where it is seen that the slopes become the same when the time 
$s$ is large enough.

\section{Conclusion and discussion}\label{section7}

In this paper, we have studied the fluctuations of the total internal energy of 
a granular gas in the stationary USF state. Using the approximation given by 
Eq. (\ref{aprox2}), it has been shown that the fluctuations of the total 
internal energy are coupled to the fluctuations of the several components of 
the total pressure tensor. The approximation is fully consistent with the 
kinetic equation for the correlation function and, in principle, is not 
limited to small inelasticities. One of the main results of the paper is the 
closed system of equations given in (\ref{ec:124}), for the first six 
moments of the 
correlation function. With them, one can calculate all the possible one-time 
correlations of the total internal energy and the different components of the 
total pressure tensor. The system depends on several complex moments of the 
one-particle distribution function in the stationary USF state. 
Since this distribution function is not known exactly, the Jenkins and 
Richman distribution has been used to $\epsilon^2$ order. For $d=2$, all the 
correlation functions have been evaluated as a function of the degree of 
inelasticity, $\epsilon$, finding a good agreement with Molecular Dynamics 
simulation results. Also, the two-time correlations have been evaluated. 

At this point it is convenient to analyze the main analogies and differences 
between the HCS and the USF state. In both cases, there is not a 
Fluctuation-Dissipation relation of the second kind, as the expression for the 
auto-correlation function of the total internal energy is not directly related 
with the coefficients of the macroscopic equation, that in this case is the 
cooling rate \cite{bgm12}. Moreover, in both the HCS state and the USF state, 
the two-time correlation function does decay as a homogeneous macroscopic 
perturbation, see Eq. (\ref{eqPsi}), so that there is a Fluctuation-Dissipation 
relation of the first kind. The main difference between the two cases resides 
in the nature of approximation given in Eq. (\ref{aprox2}). While in the HCS 
case, the 
approximate eigenfunction can be identified looking at the linearized 
homogeneous hydrodynamic equations, in the USF case the equations for the 
pressure tensor components are needed. Although, in principle, this fact does 
not have direct 
consequences at the level of macroscopic hydrodynamics, it is important at the 
level of the fluctuations. Actually, the correlation 
function, $\langle\delta\mathcal{P}_{xy}(t)\delta\mathcal{P}_{xy}(0)\rangle$, 
does not decay as a pure kinetic mode as is the case in the HCS and as was 
assumed in \cite{bgm12} and, then, the fluctuating quantity 
$\delta\mathcal{P}_{xy}(t)$ can not be treated simply as a noise in a 
consistent way (one of the conditions for the results of \cite{bgm12} to hold 
was that the correlation function of the noise decay faster than the one of the 
energy). Let us stress that, as the hydrodynamic part of the correlation 
function is of $\epsilon^2$ order, the coupling disappears in the elastic 
limit where we exactly recover the result of \cite{bgm12}. 

Finally, let us mention that many of the general properties shown in the paper 
can appear in any system beyond Navier-Stokes. Moreover, these results present 
the starting point for the complete study of the hydrodynamic fluctuating 
fields in the USF state.

\section{Acknowledgments}

This research was supported by the Ministerio de Educaci\'{o}n y
Ciencia (Spain) through Grant No. FIS2011-24460 (partially financed
by FEDER funds). 

\appendix
\section{Evolution equation for the temperature}\label{apendiceA}

The objective of this appendix is to identify the mode that emerges after a 
homogeneous perturbation of the density and the temperature. Assuming that the 
hydrodynamic stage has been reached and the distribution function is the one of 
the time-dependent USF state, we have
\begin{equation}
\frac{dT_H(t)}{dt}=-\frac{2a}{dn_H}P_{xy, H}(t)-\zeta_H(t)T_H(t)
\end{equation}
where the pressure tensor and cooling rate can be written as
\begin{eqnarray}
P_{xy,H}(t)&=&\frac{1}{2}n_Hmv_H(t)^2\tilde{P}_{xy}(\tilde{a}),\\
\zeta_H(t)&=&\frac{v_H(t)}{\lambda}\tilde{\zeta}(\tilde{a}), 
\end{eqnarray}
where $\tilde{P}_{xy}(\tilde{a})$ and $\tilde{\zeta}(\tilde{a})$ are defined in 
Eqs. (\ref{pxys}) and (\ref{zetas}) respectively. 

The deviation 
\begin{equation}
\delta\left[\frac{P_{xy,H}}{n_H}\right]
\equiv\frac{P_{xy,H}}{n_H}-\frac{P_{xy,s}}{n_s}, 
\end{equation}
around the stationary value given by $n_s$ and $T_s$ is, to linear order, 
\begin{equation}\label{ap1}
\delta\left[\frac{P_{xy,H}}{n_H}\right]=\frac{1}{2}mv_s^2\tilde{P}_{xy, s}\theta
-\frac{1}{2}mv_s^2\tilde{a}_s\frac{\partial\tilde{P}_{xy}}{\partial\tilde{a}}
(\tilde{a}_s)\left(\frac{1}{2}\theta+\rho\right), 
\end{equation}
where we have used that
\begin{equation}
\left[\frac{\partial\tilde{a}}{\partial v_H}\right]_{v_H=v_s}
=-\frac{\tilde{a}_s}{v_s}, \quad 
\left[\frac{\partial\tilde{a}}{\partial n_H}\right]_{v_H=v_s}
=-\frac{\tilde{a}_s}{n_s}. 
\end{equation}
Analogously, for the cooling rate term we have
\begin{equation}\label{ap2}
\delta[\zeta_H(t)T_H(t)]=T_s\frac{v_s}{\lambda}\left[\frac{3}{2}
\tilde{\zeta}(\tilde{a}_s)-\frac{\tilde{a}_s}{2}
\frac{d\tilde{\zeta}}{d\tilde{a}}(\tilde{a}_s)\right]\theta
+T_s\frac{v_s}{\lambda}\left[\tilde{\zeta}(\tilde{a}_s)
-\tilde{a}_s\frac{d\tilde{\zeta}}{d\tilde{a}}(\tilde{a}_s)\right]\rho. 
\end{equation}
Taking into account Eqs. (\ref{ap1}) and (\ref{ap2}), we obtain
\begin{eqnarray}
&&\frac{d\theta}{ds}=\left[2\frac{\tilde{a}_s^2}{d}
\frac{\partial\tilde{P}_{xy}}{\partial\tilde{a}}(\tilde{a}_s)-\tilde{\zeta}_s
+\tilde{a}_s\frac{d\tilde{\zeta}}{d\tilde{a}}(\tilde{a}_s)\right]\rho
\nonumber\\
&&+\left[\frac{\tilde{a}_s^2}{d}
\frac{\partial\tilde{P}_{xy}}{\partial\tilde{a}}(\tilde{a}_s)
-\frac{2\tilde{a}_s}{d}\tilde{P}_{xy,s}-\frac{3}{2}\tilde{\zeta}_s
+\frac{1}{2}\tilde{a}_s\frac{d\tilde{\zeta}}{d\tilde{a}}(\tilde{a}_s)\right]
\theta.  
\end{eqnarray}
But, as in the stationary state we have
\begin{equation}
\frac{2\tilde{a}_s}{d}\tilde{P}_{xy, s}=-\tilde{\zeta}_s, 
\end{equation}
we obtain the result of the main text, Eq. (\ref{ectheta}). 



\section{Consistency of the approximation given by Eq. 
(\ref{aprox2})}\label{apendiceB}

In this Appendix we prove that the approximation given by Eq. (\ref{aprox2}) is 
consistent with the equation for the correlation function, Eq. (\ref{eqphi}). 
Taking the scalar product with $\bar{\xi}_j(\mathbf{c}_1)$ in Eq. (\ref{eqphi}) 
and performing the approximation (\ref{aprox2}), it is obtained
\begin{equation}\label{c1}
\lambda_j\langle\bar{\xi}_j(\mathbf{c}_1)|\phi(\mathbf{c}_1,\mathbf{c}_2)\rangle
=\langle\bar{\xi}_j(\mathbf{c}_1)|\widetilde{T}_0(\mathbf{c}_1,\mathbf{c}_2)
\chi(\mathbf{c}_1)\chi(\mathbf{c}_2)\rangle, 
\end{equation}
whose consistency can be proved. Here we will use a different approach. Eq. 
(\ref{c1}) can be obtained by first integrating with respect to $\mathbf{c}_2$ in Eq. (\ref{eqphi}) 
and then take the scalar product with $\bar{\xi}_j(\mathbf{c}_1)$ in the 
$\mathbf{c}_1$ space. Once the first step is done, it is obtained
\begin{equation}\label{beLambda}
\Lambda(\mathbf{c})\chi(\mathbf{c})
=-\tilde{a}_sc_y\frac{\partial}{\partial c_x}\chi(\mathbf{c}), 
\end{equation}
where we have used that
\begin{equation}
\int d\mathbf{c}_2\phi(\mathbf{c}_1,\mathbf{c}_2)=-\chi(\mathbf{c}). 
\end{equation}
Note that Eq. (\ref{beLambda}) is nothing but the non-linear Boltzmann 
equation for the stationary state, but expressed in terms of the linearized 
Boltzmann operator. And then, approximation (\ref{aprox2}) has to be consistent 
when applied to Eq. (\ref{beLambda}), so that what we have to prove is 
\begin{equation}\label{apBE}
\lambda_i\langle\bar{\xi}_i(\mathbf{c})|\chi(\mathbf{c})\rangle
=\tilde{a}_s\langle\bar{\xi}_i(\mathbf{c})|
c_y\frac{\partial}{\partial c_x}\chi(\mathbf{c})\rangle, \quad i=2,3,4. 
\end{equation}
For $i=1$ the previous equation trivially holds. 

Eqs. (\ref{apBE}) can be expressed in a different basis of the subspace 
$\{\bar{\xi}_i\}_{i=1}^4$. It turns useful to use the basis $\{h_i\}_{i=1}^4$, 
where
\begin{equation}
\mathbf{h}(\mathbf{c})=
\left[\begin{array}{c} 1 \\ \frac{2c^2}{d}-1 \\ 2c_xc_y\\ 2c_y^2\\ 
\end{array}\right], 
\end{equation}
because we have
\begin{equation}
\int d\mathbf{c}\mathbf{h}(\mathbf{c})\delta\chi(\mathbf{c},s)
=\left[\begin{array}{c} \rho \\ \theta(s) \\ \Pi_{xy}(s) \\ \Pi_{yy}(s)\\ 
\end{array}\right], 
\end{equation}
and then, Eqs. (\ref{apBE}) can be written as
\begin{equation}\label{ecM}
M\left[\begin{array}{c} 1 \\ 0 \\ \tilde{P}_{xy, s}\\ \tilde{P}_{yy, s}\\ 
\end{array}\right]
=-\tilde{a}_s\left[\begin{array}{c} 0 \\ \frac{2}{d}\tilde{P}_{xy, s}\\ 
\tilde{P}_{yy, s}\\ 0 \\ 
\end{array}\right], 
\end{equation}
where we have used that 
\begin{equation}
\int d\mathbf{c}\mathbf{h}(\mathbf{c})\chi(\mathbf{c})
=\left[\begin{array}{c} 1 \\ 0 \\ \tilde{P}_{xy, s}\\ \tilde{P}_{yy, s}\\ 
\end{array}\right], \quad
\int d\mathbf{c}\mathbf{h}(\mathbf{c})\chi(\mathbf{c})
c_y\frac{\partial}{\partial c_x}\chi(\mathbf{c})
=-\left[\begin{array}{c} 0 \\ \frac{2}{d}\tilde{P}_{xy, s}\\ 
\tilde{P}_{yy, s}\\ 0 \\ 
\end{array}\right].
\end{equation}
Taking the explicit expressions of $M$, $\tilde{P}_{xy, s}$, and 
$\tilde{P}_{yy, s}$ as a function of $\tilde{\zeta}_s$ and $\beta$ of the main 
text, it is straightforward to prove the validity of Eq. (\ref{ecM}). 

\section{Evaluation of the $C_{ij}$ coefficients}\label{apendiceC}

In this Appendix we calculate the coefficients $C_{ij}$ defined in Eq. (\ref{defCij}), starting from Eq. (\ref{ec:124}). As said, we only have $6$ independent equations, because of the relation 
between the third and fourth mode. The corresponding equation to $i=j=2$ is
\begin{equation}\label{eq1}
2\lambda_2\langle\bar{\xi}_2(\mathbf{c}_1)\bar{\xi}_2(\mathbf{c}_2)|
\phi(\mathbf{c}_1,\mathbf{c}_2)\rangle=
\langle\bar{\xi}_2(\mathbf{c}_1)\bar{\xi}_2(\mathbf{c}_2)|
\widetilde{T}_0(\mathbf{c}_1,\mathbf{c}_2)
\chi(\mathbf{c}_1)\chi(\mathbf{c}_2)\rangle.  
\end{equation}
For $i=2$ and $j=3$ we have two equations, one associated to the real part 
\begin{eqnarray}\label{eq2}
(\lambda_2+\lambda_3^R)
\langle\bar{\xi}_2(\mathbf{c}_1)\bar{\xi}_3^R(\mathbf{c}_2)|
\phi(\mathbf{c}_1,\mathbf{c}_2)\rangle
-\lambda_3^I\langle\bar{\xi}_2(\mathbf{c}_1)\bar{\xi}_3^I(\mathbf{c}_2)|
\phi(\mathbf{c}_1,\mathbf{c}_2)\rangle \nonumber\\
=\langle\bar{\xi}_2(\mathbf{c}_1)\bar{\xi}_3^R(\mathbf{c}_2)|
\widetilde{T}_0(\mathbf{c}_1,\mathbf{c}_2)
\chi(\mathbf{c}_1)\chi(\mathbf{c}_2)\rangle,  
\end{eqnarray}
and other to the imaginary part
\begin{eqnarray}\label{eq3}
(\lambda_2+\lambda_3^R)
\langle\bar{\xi}_2(\mathbf{c}_1)\bar{\xi}_3^I(\mathbf{c}_2)|
\phi(\mathbf{c}_1,\mathbf{c}_2)\rangle
-\lambda_3^I\langle\bar{\xi}_2(\mathbf{c}_1)\bar{\xi}_3^R(\mathbf{c}_2)|
\phi(\mathbf{c}_1,\mathbf{c}_2)\rangle \nonumber\\
=\langle\bar{\xi}_2(\mathbf{c}_1)\bar{\xi}_3^I(\mathbf{c}_2)|
\widetilde{T}_0(\mathbf{c}_1,\mathbf{c}_2)
\chi(\mathbf{c}_1)\chi(\mathbf{c}_2)\rangle, 
\end{eqnarray}
where we have used the decomposition into the real and imaginary part of the 
third eigenvalue and eigenfunctions given by Eqs. (\ref{lri})-(\ref{xiri}). For 
$i=j=3$ we also have two independent equations
\begin{eqnarray}\label{eq4}
2\lambda_3^R[\langle\bar{\xi}_3^R(\mathbf{c}_1)\bar{\xi}_3^R(\mathbf{c}_2)|
\phi(\mathbf{c}_1,\mathbf{c}_2)\rangle-
\langle\bar{\xi}_3^I(\mathbf{c}_1)\bar{\xi}_3^I(\mathbf{c}_2)|
\phi(\mathbf{c}_1,\mathbf{c}_2)\rangle]\nonumber\\
-4\lambda_3^I\langle\bar{\xi}_3^R(\mathbf{c}_1)\bar{\xi}_3^I(\mathbf{c}_2)|
\phi(\mathbf{c}_1,\mathbf{c}_2)\rangle\nonumber\\
=\langle\bar{\xi}_3^R(\mathbf{c}_1)\bar{\xi}_3^R(\mathbf{c}_2)|
\widetilde{T}_0(\mathbf{c}_1,\mathbf{c}_2)
\chi(\mathbf{c}_1)\chi(\mathbf{c}_2)\rangle
-\langle\bar{\xi}_3^I(\mathbf{c}_1)\bar{\xi}_3^I(\mathbf{c}_2)|
\widetilde{T}_0(\mathbf{c}_1,\mathbf{c}_2)
\chi(\mathbf{c}_1)\chi(\mathbf{c}_2)\rangle, 
\end{eqnarray}
and
\begin{eqnarray}\label{eq5}
\lambda_3^I[\langle\bar{\xi}_3^R(\mathbf{c}_1)\bar{\xi}_3^R(\mathbf{c}_2)|
\phi(\mathbf{c}_1,\mathbf{c}_2)\rangle
-\langle\bar{\xi}_3^I(\mathbf{c}_1)\bar{\xi}_3^I(\mathbf{c}_2)|
\phi(\mathbf{c}_1,\mathbf{c}_2)\rangle]\nonumber\\
+2\lambda_3^R\langle\bar{\xi}_3^R(\mathbf{c}_1)\bar{\xi}_3^I(\mathbf{c}_2)|
\phi(\mathbf{c}_1,\mathbf{c}_2)\rangle
=\langle\bar{\xi}_3^R(\mathbf{c}_1)\bar{\xi}_3^I(\mathbf{c}_2)|
\widetilde{T}_0(\mathbf{c}_1,\mathbf{c}_2)
\chi(\mathbf{c}_1)\chi(\mathbf{c}_2)\rangle. 
\end{eqnarray}
Finally, there is an additional equation corresponding to $i=3$, $j=4$
\begin{eqnarray}\label{eq6}
2\lambda_3^R\langle\bar{\xi}_3^R(\mathbf{c}_1)\bar{\xi}_3^R(\mathbf{c}_2)|
\phi(\mathbf{c}_1,\mathbf{c}_2)\rangle
+\langle\bar{\xi}_3^I(\mathbf{c}_1)\bar{\xi}_3^I(\mathbf{c}_2)|
\phi(\mathbf{c}_1,\mathbf{c}_2)\rangle]\nonumber\\
=\langle\bar{\xi}_3^R(\mathbf{c}_1)\bar{\xi}_3^R(\mathbf{c}_2)|
\widetilde{T}_0(\mathbf{c}_1,\mathbf{c}_2)
\chi(\mathbf{c}_1)\chi(\mathbf{c}_2)\rangle
+\langle\bar{\xi}_3^I(\mathbf{c}_1)\bar{\xi}_3^I(\mathbf{c}_2)|
\widetilde{T}_0(\mathbf{c}_1,\mathbf{c}_2)
\chi(\mathbf{c}_1)\chi(\mathbf{c}_2)\rangle,
\end{eqnarray}
that can be written in terms of the third mode because 
$\lambda_4=\lambda_3^*$ and $\bar{\xi}_4=\bar{\xi}_3^*$. 

The scalar products 
$\langle\bar{\xi}_i(\mathbf{c}_1)\bar{\xi}_j(\mathbf{c}_2)|
\phi(\mathbf{c}_1,\mathbf{c}_2)\rangle$ can be written in terms of the 
$C_{ij}$ coefficients through Eq. (\ref{ecxixi}), so that
the system of equations (\ref{eq1})-(\ref{eq6}) is a linear system of six 
equations for the six unknown coefficients 
$\{C_{22},C_{23},C_{24},C_{33},C_{34},C_{44}\}$. Let us note that, until 
now, the results are valid for any dimension, $d$, and the only approximation 
made was the one given by Eq. (\ref{aprox2}). Of course, it still 
remains to evaluate the coefficients
\begin{equation}\label{ectij} 
\langle\bar{\xi}_i(\mathbf{c}_1)\bar{\xi}_j(\mathbf{c}_2)|
\widetilde{T}_0(\mathbf{c}_1,\mathbf{c}_2)
\chi(\mathbf{c}_1)\chi(\mathbf{c}_2)\rangle=\sum_{l=1}^4\xi_{il}\xi_{jl}T_{ll}
+\sum_{k>l=1}^4(\xi_{ik}\xi_{jl}+\xi_{il}\xi_{jk})T_{kl}, 
\end{equation}
where we have introduced the matrix elements
\begin{equation}\label{tij}
T_{ij}=\int d\mathbf{c}_1\int d\mathbf{c}_2b_i(\mathbf{c}_1)b_j(\mathbf{c}_2)
\widetilde{T}_0(\mathbf{c}_1,\mathbf{c}_2)\chi(\mathbf{c}_1)\chi(\mathbf{c}_2)
= \int d\mathbf{c}_1\int d\mathbf{c}_2\chi(\mathbf{c}_1)\chi(\mathbf{c}_2)
T_0(\mathbf{c}_1,\mathbf{c}_2)b_i(\mathbf{c}_1)b_j(\mathbf{c}_2), 
\end{equation}
with
\begin{equation}
T_0(\mathbf{c}_1,\mathbf{c}_2)=\int d\sig\Theta(\mathbf{c}_{12}\cdot\sig)
(\mathbf{c}_{12}\cdot\sig)[b_{\boldsymbol{\sigma}}(1,2)-1]. 
\end{equation}


The first coefficients, $\{T_{1j}\}_{j=1}^4$, can be easily calculated. 
In effect, 
\begin{equation}
T_{11}=0, 
\end{equation}
due to the conservation of the total number of particles and the second is 
related with the cooling rate
\begin{equation}
T_{12}=-\frac{d}{2}\tilde{\zeta}_s, 
\end{equation}
by Eq. (\ref{zetas}). On the other hand, taking into account the equation for 
$\chi$, Eq. (\ref{beLambda}), we have
\begin{equation}
T_{13}=
-\tilde{a}_s\int d\mathbf{c}c_{x}c_{y}^2\frac{\partial}{\partial c_x}
\chi(\mathbf{c})=\frac{1}{2}\tilde{a}_s\tilde{P}_{yy, s}, 
\end{equation}
and 
\begin{equation}
T_{14}=-\tilde{a}_s\int d\mathbf{c}c_{y}^2\frac{\partial}{\partial c_x}
\chi(\mathbf{c})=0. 
\end{equation}

To evaluate the other coefficients we have to calculate explicitly 
$T_0(\mathbf{c}_1,\mathbf{c}_2)b_i(\mathbf{c}_1)b_j(\mathbf{c}_2)$. In fact, in 
reference \cite{bgmr04} the term 
$T_0(\mathbf{c}_1,\mathbf{c}_2)c_1^2c_2^2$ was already calculated obtaining
\begin{eqnarray}
T_0(\mathbf{c}_1,\mathbf{c}_2)c_1^2c_2^2
=-\frac{\pi^{(d-1)/2}}{\Gamma\left(\frac{d+5}{2}\right)}
\left[\frac{(1-\alpha^2)(d+1+2\alpha^2)}{16}g^5\right.\nonumber\\
\left.+\frac{d+5-\alpha^2(d+1)+4\alpha}{4}g^3G^2
-\frac{1+\alpha}{2}(2d+3-3\alpha)g(\mathbf{g}\cdot\mathbf{G})^2\right],  
\end{eqnarray}
where we have introduced the new variables
\begin{eqnarray}
\mathbf{g}&=&\mathbf{c}_1-\mathbf{c}_2, \\
\mathbf{G}&=&\frac{1}{2}(\mathbf{c}_1+\mathbf{c}_2). 
\end{eqnarray}
For the rest of coefficients, we first evaluate 
$[b_{\boldsymbol{\sigma}}(1,2)-1]b_i(\mathbf{c}_1)b_j(\mathbf{c}_2)$. Using the 
collision rule, Eq. (\ref{collisionRule}), it is obtained 
\begin{eqnarray}
[b_{\boldsymbol{\sigma}}(1,2)-1]c_1^2c_{2x}c_{2y}=\frac{1+\alpha}{2}
(\sig\cdot\mathbf{g})[c_1^2(c_{2x}\hat{\sigma}_y+c_{2y}\hat{\sigma}_x)
-2(\sig\cdot\mathbf{c}_1)c_{2x}c_{2y}]
\nonumber\\
+\frac{(1+\alpha)^2}{4}
(\sig\cdot\mathbf{g})^2[c_1^2\hat{\sigma}_x\hat{\sigma}_y+c_{2x}c_{2y}
-2(\sig\cdot\mathbf{c}_1)(c_{2x}\hat{\sigma}_y+c_{2y}\hat{\sigma}_x)]
\nonumber\\
+\frac{(1+\alpha)^3}{8}
(\sig\cdot\mathbf{g})^3[c_{2x}\hat{\sigma}_y+c_{2y}\hat{\sigma}_x
-2(\sig\cdot\mathbf{c}_1)\hat{\sigma}_x\hat{\sigma}_y]
+\frac{(1+\alpha)^4}{16}
(\sig\cdot\mathbf{g})^4\hat{\sigma}_x\hat{\sigma}_y, \nonumber\\
\end{eqnarray}
\begin{eqnarray}
[b_{\boldsymbol{\sigma}}(1,2)-1]c_1^2c_{2y}^2
=(1+\alpha)(\sig\cdot\mathbf{g})[c_1^2c_{2y}\hat{\sigma}_y
-c_{2y}^2(\sig\cdot\mathbf{c}_1)]
\nonumber\\
+\frac{(1+\alpha)^2}{4}(\sig\cdot\mathbf{g})^2
[c_1^2\hat{\sigma}_y^2+c_{2y}^2-4(\sig\cdot\mathbf{c}_1)c_{2y}\hat{\sigma}_y]
\nonumber\\
+\frac{(1+\alpha)^3}{4}(\sig\cdot\mathbf{g})^3
[c_{2y}\hat{\sigma}_y-(\sig\cdot\mathbf{c}_1)\hat{\sigma}_y^2]
+\frac{(1+\alpha)^4}{16}
(\sig\cdot\mathbf{g})^4\hat{\sigma}_y^2, 
\end{eqnarray}
\begin{eqnarray}
[b_{\boldsymbol{\sigma}}(1,2)-1]c_{1x}c_{1y}c_{2x}c_{2y}=
\frac{1+\alpha}{2}
(\sig\cdot\mathbf{g})(\hat{\sigma}_yc_{1x}c_{2x}g_y+\hat{\sigma}_xc_{1y}c_{2y}g_x)
\nonumber\\
+\frac{(1+\alpha)^2}{4}
(\sig\cdot\mathbf{g})^2[\hat{\sigma}_x\hat{\sigma}_yg_xg_y
-\hat{\sigma}_x^2c_{1y}c_{2y}-\hat{\sigma}_y^2c_{1x}c_{2x}]
\nonumber\\
-\frac{(1+\alpha)^3}{8}
(\sig\cdot\mathbf{g})^3(\hat{\sigma}_x\hat{\sigma}_y^2g_x
+\hat{\sigma}_x^2\hat{\sigma}_yg_y)
-\frac{(1+\alpha)^4}{16}(\sig\cdot\mathbf{g})^4\hat{\sigma}_x^2\hat{\sigma}_y^2, 
\end{eqnarray}
\begin{eqnarray}
[b_{\boldsymbol{\sigma}}(1,2)-1]c_{1x}c_{1y}c_{2y}^2=
\frac{1+\alpha}{2}(\sig\cdot\mathbf{g})
[\hat{\sigma}_y(2c_{1x}c_{1y}c_{2y}-c_{1x}c_{2y}^2)-\hat{\sigma}_xc_{1y}c_{2y}^2]
\nonumber\\
+\frac{(1+\alpha)^2}{4}
(\sig\cdot\mathbf{g})^2[\hat{\sigma}_y^2(c_{1x}c_{1y}-2c_{1x}c_{2y})
+\hat{\sigma}_x\hat{\sigma}_y(c_{2y}^2-2c_{1y}c_{2y})]
\nonumber\\
+\frac{(1+\alpha)^3}{8}
(\sig\cdot\mathbf{g})^3[\hat{\sigma}_x\hat{\sigma}_y^2(2c_{2y}-c_{1y})
-\hat{\sigma}_y^3c_{1x}]
+\frac{(1+\alpha)^4}{16}(\sig\cdot\mathbf{g})^4\hat{\sigma}_x\hat{\sigma}_y^3, 
\nonumber\\
\end{eqnarray}
and 
\begin{eqnarray}
[b_{\boldsymbol{\sigma}}(1,2)-1]c_{1y}^2c_{2y}^2=
(1+\alpha)(\sig\cdot\mathbf{g})\hat{\sigma}_y(c_{1y}^2c_{2y}-c_{1y}c_{2y}^2)
\nonumber\\
+\frac{(1+\alpha)^2}{4}
(\sig\cdot\mathbf{g})^2\hat{\sigma}_y^2(c_{1y}^2+c_{2y}^2-4c_{1y}c_{2y})
\nonumber\\
-\frac{(1+\alpha)^3}{4}
(\sig\cdot\mathbf{g})^3\hat{\sigma}_y^3g_y
+\frac{(1+\alpha)^4}{16}(\sig\cdot\mathbf{g})^4\hat{\sigma}_y^4. 
\end{eqnarray}
After multiplying by $\sig\cdot\mathbf{g}$, the $\sig$-integrals can be 
calculated with the aid of 
\begin{equation}
\int d\sig\Theta(\sig\cdot\mathbf{g})(\sig\cdot\mathbf{g})^2\hat{\sigma}_i
=\frac{\pi^{\frac{d-1}{2}}}{\Gamma\left(\frac{d+3}{2}\right)}gg_i, 
\end{equation}
\begin{equation}
\int d\sig\Theta(\sig\cdot\mathbf{g})(\sig\cdot\mathbf{g})^3
=\frac{\pi^{\frac{d-1}{2}}}{\Gamma\left(\frac{d+3}{2}\right)}g^3, 
\end{equation}
\begin{equation}
\int d\sig\Theta(\sig\cdot\mathbf{g})(\sig\cdot\mathbf{g})^3
\hat{\sigma}_i\hat{\sigma}_j
=\frac{\pi^{\frac{d-1}{2}}}{2\Gamma\left(\frac{d+5}{2}\right)}
(3gg_ig_j+g^3\delta_{ij}),
\end{equation}
\begin{equation}
\int d\sig\Theta(\sig\cdot\mathbf{g})(\sig\cdot\mathbf{g})^4\hat{\sigma}_i
=\frac{2\pi^{\frac{d-1}{2}}}{\Gamma\left(\frac{d+5}{2}\right)}g^3g_i, 
\end{equation}
\begin{equation}
\int d\sig\Theta(\sig\cdot\mathbf{g})(\sig\cdot\mathbf{g})^4
\hat{\sigma}_y^2\hat{\sigma}_j
=\frac{\pi^{\frac{d-1}{2}}}{\Gamma\left(\frac{d+7}{2}\right)}[3gg_y^2g_j+g^3g_j
+2g^3g_y\delta_{yj}], 
\end{equation}
\begin{equation}
\int d\sig\Theta(\sig\cdot\mathbf{g})(\sig\cdot\mathbf{g})^4
\hat{\sigma}_x\hat{\sigma}_y\hat{\sigma}_z
=\frac{3\pi^{\frac{d-1}{2}}}{\Gamma\left(\frac{d+7}{2}\right)}gg_xg_yg_z, 
\end{equation}
\begin{equation}
\int d\sig\Theta(\sig\cdot\mathbf{g})(\sig\cdot\mathbf{g})^5
\hat{\sigma}_i\hat{\sigma}_j
=\frac{\pi^{\frac{d-1}{2}}}{\Gamma\left(\frac{d+7}{2}\right)}
(5g^3g_ig_j+g^5\delta_{ij}),
\end{equation}
\begin{equation}
\int d\sig\Theta(\sig\cdot\mathbf{g})(\sig\cdot\mathbf{g})^5
\hat{\sigma}_y^3\hat{\sigma}_j
=\frac{3\pi^{\frac{d-1}{2}}}{2\Gamma\left(\frac{d+9}{2}\right)}
[5gg_y(g^2+g_y^2)g_j+g^3(g^2+5g_y^2)\delta_{yj}],
\end{equation}
calculated for arbitrary dimension and 
\begin{equation}
\int d\sig\Theta(\sig\cdot\mathbf{g})(\sig\cdot\mathbf{g})^5
\hat{\sigma}_x^2\hat{\sigma}_j^2
=\frac{3\pi^{\frac{1}{2}}}{2\Gamma\left(\frac{11}{2}\right)}
[g^3(g^2+g_x^2)+8g^3g_xg_j\delta_{xj}+(g^3+5gg_x^2)g_j^2],
\end{equation}
for $d=2$. 

In the following we will restrict ourselves to the case $d=2$ and will use the 
Jenkins and Richman distribution to order $\epsilon^2$ \cite{jr88}
\begin{equation}
\chi(\mathbf{c})\approx\frac{e^{-c^2}}{\pi}\left[1-\epsilon\sqrt{2}c_xc_y
+\epsilon^2\left(\frac{1}{4}-c_y^2+c_x^2c_y^2\right)\right]. 
\end{equation}
To the same order, we have
\begin{eqnarray}
\chi(\mathbf{c}_1)\chi(\mathbf{c}_2)\approx\frac{1}{\pi^2}e^{-c_1^2-c_2^2}
\left[1-\epsilon\sqrt{2}(c_{1x}c_{1y}+c_{2x}c_{2y})\right.\nonumber\\
\left.+\epsilon^2\left(\frac{1}{4}-c_{1y}^2-c_{2y}^2+c_{1x}^2c_{1y}^2
+c_{2x}^2c_{2y}^2+2c_{1x}c_{1y}c_{2x}c_{2y}\right)\right], 
\end{eqnarray}
or, in terms of the new variables $\{\mathbf{g},\mathbf{G}\}$
\begin{eqnarray}\label{chichi}
\chi(\mathbf{c}_1)\chi(\mathbf{c}_2)\approx\frac{e^{-\frac{1}{2}g^2-2G^2}}{\pi^2}
\left\{1-\frac{\epsilon}{\sqrt{2}}(g_{x}g_{y}+4G_{x}G_{y})\right.\nonumber\\
\left.+\frac{\epsilon^2}{4}\left[2+(g_x^2-2)g_y^2+8g_xg_yG_xG_y+8(2G_x^2-1)G_y^2
\right]\right\}. 
\end{eqnarray}
The velocity integrals given by Eq. (\ref{tij}) can be calculated with 
the aid of Mathematica, obtaining to $\epsilon^2$ order
\begin{eqnarray}
T_{22}&=&-\frac{3}{2}\sqrt{\frac{\pi}{2}}\epsilon^2, \\
T_{23}&=&\frac{5}{8}\sqrt{\pi}\epsilon, \\
T_{24}&=&-\frac{1}{8}\sqrt{\frac{\pi}{2}}\epsilon^2, \\
T_{33}&=&-\frac{19}{64}\sqrt{\frac{\pi}{2}}\epsilon^2, \\
T_{34}&=&\frac{5}{16}\sqrt{\pi}\epsilon, \\
T_{44}&=&\frac{11}{64}\sqrt{\frac{\pi}{2}}\epsilon^2. 
\end{eqnarray}

Finally, by substituting the obtained expressions of the $T_{ij}$ coefficients 
into Eq. (\ref{ectij}) and that into Eqs. (\ref{eq1})-(\ref{eq6}), we obtain 
the above mentioned linear system for $C_{ij}$. This system is solved with the 
aid of Mathematica obtaining the expressions of the main text.

\end{document}